\documentclass[aps,prd,reprint,longbibliography,nofootinbib,superscriptaddress,floatfix]{revtex4-2}

\usepackage{mathrsfs}
\usepackage{comment}
\usepackage[T1]{fontenc}
\usepackage{lmodern}
\usepackage[utf8]{inputenc}
\usepackage{amsmath,amssymb,bm}
\usepackage{graphicx}
\usepackage{xcolor}
\definecolor{myblue}{RGB}{71,81,200}
\usepackage{hyperref}
\hypersetup{colorlinks=true,linkcolor=myblue,citecolor=myblue,urlcolor=myblue}
\newcommand{\Rew}{\operatorname{Re}}
\newcommand{\Imw}{\operatorname{Im}}

\newcommand{\dd}{\mathrm{d}}

\newcommand{\GEepA}{0.3041165140479288814}
\newcommand{\GEepQ}{0.9523626647492863734}
\newcommand{\GEepOmegaRe}{0.5643982322682300042}
\newcommand{\GEepOmegaIm}{0.0283045487334943083}
\newcommand{\GEapproachT}{10^{-12}}
\newcommand{\GEpairGap}{2.6179824601\times10^{-9}}
\newcommand{\GEwideSamples}{433}
\newcommand{\GElocalSamples}{197}

\newcommand{\ScalarEPa}{0.791531800866223}
\newcommand{\ScalarEPQ}{0.610458783564616}
\newcommand{\ScalarEPOmegaRe}{0.492968350630978}
\newcommand{\ScalarEPOmegaIm}{0.0253077532073792}
\newcommand{\ScalarApproachT}{10^{-12}}
\newcommand{\ScalarPairGap}{8.3669819433\times10^{-10}}
\newcommand{\ScalarBaseSamples}{2911}
\newcommand{\ScalarWideSamples}{2402}
\newcommand{\ScalarLocalSamples}{1247}
\begin{document}
\title{Quasinormal-mode encounters: Pseudospectra near Kerr--Newman exceptional points}
\author{Kyriakos Destounis}
\affiliation{CENTRA, Departamento de F\'isica, Instituto Superior T\'ecnico -- IST, Universidade de Lisboa -- UL, Avenida Rovisco Pais 1, 1049-001 Lisboa, Portugal}
\affiliation{School of Applied Mathematical and Physical Sciences -- SEMFE, Physics Department, National Technical University of Athens -- Polytechnic Campus, Zografou 15772, Athens, Greece}
%
\begin{abstract}
Exceptional points (EPs), at which two quasinormal frequencies and their eigenvectors coalesce, can enhance the spectral instability of black hole quasinormal modes. We investigate this behavior in Kerr--Newman spacetimes through pseudospectra and condition numbers for neutral massless scalar fields and coupled gravitoelectromagnetic (GE) perturbations. For the scalar field, we follow the $(\ell,m)=(1,1)$ spheroidal branch and derive the energy norm from the stress-energy tensor. For GE perturbations, we use a coupled four-field system equipped with a positive radiative energy. Along matched parameter paths toward the selected EPs, the frequency splitting follows a square-root law and the condition numbers grow as the inverse square root of the distance to the EP. Their product, which approximates the norm of the double-pole coefficient, does not depend on the offset or, for GE, on the angular truncation and projection used. The pseudospectral contours resolve the transition from individual-mode sensitivity to the collective response of the pair. At the selected backgrounds and resolution, both pair members are more sensitive than two more-damped modes at the same background, but this ordering holds only below a crossover offset that decreases with radial resolution. For the GE pair, angular refinement also displaces the coalescence by more than the selected offset. Within the discretized problems and energy norms considered here, these results describe qualitative trends of spectral instability near EPs rather than continuum magnitudes, and motivate further study of their possible role in black-hole ringdown.
\end{abstract}

\maketitle

\section{Introduction}
\label{sec:introduction}

Quasinormal modes (QNMs) describe the resonant response of a black hole (BH) to perturbations~\cite{Kokkotas:1999bd,Berti:2009kk,Konoplya:2011qq}. The real and imaginary parts of their frequencies determine the oscillation and damping rates, which depend on the remnant's mass, spin and electric charge. Measuring these frequencies is the basis of BH spectroscopy, through which one can infer remnant properties and test general relativity~\cite{Dreyer2004,Carullo:2025oms,Berti:2025hly,Franchini:2023eda,Berti:2015itd}. A QNM frequency alone does not determine the excitation of the mode or its sensitivity to changes in the evolution equations. These questions require information beyond the spectrum, supplied by tools such as pseudospectra, condition numbers and transient-growth analysis~\cite{TrefethenEmbree2005,Jaramillo2021,Destounis:2021lum,Boyanov:2022ark,Boyanov:2023qqf,Destounis:2023nmb,Cai2025,Destounis:2023ruj,Jaramillo:2022kuv,Carballo:2025ajx,Besson:2025ghu}. Studies of several BH systems have found QNM shifts that are disproportionately large compared with the perturbations of the equations that produce them~\cite{Jaramillo:2021tmt,Cheung:2021bol,Berti:2022xfj,Cardoso:2024mrw,Destounis:2025dck,Courty:2023rxk,Torres:2023nqg,Rosato:2024arw,Yang:2024vor,Bizon:2026dtx,May:2026qpf}.

The structure of the BH response extends beyond ordinary QNM resonances, exhibiting algebraically special solutions~\cite{Berti:2009kk,Dias:2013hn}, late-time tails~\cite{Price1972,Brady:1996za,Jansen:2017oag,Cardoso:2017soq,Destounis:2019hca,Destounis:2020pjk}, direct waves~\cite{Leaver1986,Lu:2025vol,Ma:2026hcb}, and exceptional structures in the QNM spectrum~\cite{Heiss:2012dx,Meng:2024,Zhang:2026inu,Cavalcante:2024swt,CavalcanteRichartz2026,MisasArcos2026,PanossoMacedo:2025xnf,Zhang:2026uoo,Morikawa:2026fzn}. An EP occurs at particular background parameters where two QNM frequencies and their respective eigenvectors coalesce. As the modes approach one another, their individual contributions can become highly sensitive even when their combined response remains finite~\cite{MotohashiResonance2025,KNOutlookOshita2025}. Relating this behavior to an observable waveform requires treating the pair, as well as its separate members~\cite{YangBertiFranchini2025}.

The coupled gravitational and electromagnetic spectrum of Kerr--Newman (KN) BHs offers a setting in which to study these effects \cite{Pombo:2026xsj,pombo_2026_21359120,Hintz2026,Destounis2026}. Earlier calculations mapped this spectrum and identified interactions between photon-sphere and near-horizon modes, together with eigenvalue repulsion~\cite{DiasGodazgarSantos2015,DiasEtAl2022,DiasGodazgarSantos2022}. Scalar Kerr and KN spectra also exhibit mode interactions and EPs~\cite{Cai2025,Davey2023,CavalcanteErgodic2026,CavalcanteRichartz2026}. Pseudospectral scaling near exceptional structures has been studied in a modified Regge--Wheeler potential~\cite{Cao2026}, and coupled GE KN EPs and mode exchange have been reported in Ref.~\cite{MisasArcos2026}.

The choice of norm and allowed perturbations is essential to a pseudospectral calculation \cite{Gasperin:2021kfv}. Even separated equations and accurate QNM frequencies~\cite{Hintz2026,Destounis2026} do not fix a response measure on their own. We use an energy obtained from the scalar stress current and a positive energy for the coupled four-field GE system, accounting for the conversion from an equation source to forcing of the evolution state. Each near-EP pair is compared with two more-damped modes at the same background. We also follow a matched GE parameter path toward coalescence, which allows the EP scaling to be tested without changing the mode identification between successive backgrounds. Because condition numbers and pseudospectra of discretized dissipative BH problems need not converge with resolution~\cite{Jaramillo2021,Destounis:2021lum}, we also vary the offset from the EP, the radial resolution and the angular truncation, and distinguish the features that persist under these changes from those that do not.

We introduce the common definitions in Sec.~\ref{sec:framework}, followed by the scalar formulation and results in Secs.~\ref{sec:scalar} and \ref{sec:results}. Sections~\ref{sec:ge}--\ref{sec:measured-conditioning} develop the coupled GE calculation. The physical interpretation, together with resolution and truncation tests, is discussed in Sec.~\ref{sec:discussion}, before the implications for ringdown and evolving backgrounds are considered in Sec.~\ref{sec:conclusions-outlook}. Derivations and numerical checks are collected in the appendices.

\section{Energy-norm resolvent and condition number}
\label{sec:framework}

We work in geometrized units $G=c=1$, and set the KN BH mass to unity, unless the mass is restored explicitly. With the Fourier convention $e^{-i\omega T+im\varphi}$, damped QNMs have $\Imw\omega<0$. Consider a frequency-independent discretized evolution matrix $L_N$, where $N$ is the number of Chebyshev-Lobatto radial collocation points. Any angular truncation $L_{\rm ang}$ will be stated separately. The energy defines a positive Hermitian matrix $G_N$ through
\begin{equation}
 G_N=C_N^\dagger C_N,\qquad
 \|u\|_E^2=u^\dagger G_Nu=\|C_Nu\|_2^2.
 \label{eq:paper-energy-coordinates}
\end{equation}
Here $C_N$ is an upper-triangular Cholesky factor of $G_N$. In the coordinates $C_Nu$, the Euclidean length of the state equals its energy norm~\cite{Jaramillo2021}.

For a complex frequency $z$ outside the spectrum, the resolvent is
\begin{equation}
 R_N(z)=(zI-L_N)^{-1}.
 \label{eq:paper-resolvent}
\end{equation}
The inverse of its largest response in the energy norm is the minimum singular value $s_{\rm min}$, i.e.,
\begin{equation}
 \sigma_E(z)=s_{\min}
 \!\left[C_N(zI-L_N)C_N^{-1}\right]
 =\|R_N(z)\|_E^{-1}.
 \label{eq:paper-minimum-residual}
\end{equation}
At an eigenfrequency this singular value vanishes. For $\epsilon>0$, the energy-norm pseudospectrum is defined by
\begin{equation}
 \sigma_E^{(\epsilon)}(L_N)=\{z\in\mathbb C:s_E(z)<\epsilon\}.
 \label{eq:paper-pseudospectrum}
\end{equation}
This set includes the spectrum, $\sigma(L_N)$, and all frequencies at which $\|R_N(z)\|_E>\epsilon^{-1}$. It also comprises the eigenvalues of $L_N+\delta L_N$ that can be reached by unrestricted matrix perturbations satisfying $\|\delta L_N\|_E<\epsilon$~\cite{TrefethenEmbree2005}. The same norm measures the response and the size of the perturbation. The threshold $\epsilon$ has the frequency units of $L_N$, rather than units of energy.

For a simple eigenvalue $\omega_j$, let $x_j$ and $y_j$ satisfy $L_Nx_j=\omega_jx_j$ and $y_j^\dagger L_N=\omega_jy_j^\dagger$. Its absolute condition number is~\cite{TrefethenEmbree2005,Jaramillo2021}
\begin{equation}
 \kappa_E(\omega_j)=
 \frac{\|C_Nx_j\|_2\,\|C_N^{-\dagger}y_j\|_2}
 {|y_j^\dagger x_j|}.
 \label{eq:paper-kappa}
\end{equation}
The norm of the left eigenvector involves the inverse energy matrix, $\|C_N^{-\dagger}y_j\|_2^2=y_j^\dagger G_N^{-1}y_j$. A perturbation of the evolution matrix satisfies the first-order bound $|\delta\omega_j|\leq\kappa_E(\omega_j)\|\delta L_N\|_E$~\cite{Alam2015}. Near an isolated simple pole, the residual has the expansion
\begin{equation}
 \sigma_E(z)=\frac{|z-\omega_j|}{\kappa_E(\omega_j)}
             +\mathcal O(|z-\omega_j|^2).
 \label{eq:paper-pole-slope}
\end{equation}
A larger condition number allows a larger local frequency displacement under a perturbation of the same size.

At a second-order EP, where two eigenvalues and eigenvectors coalesce (for higher-order EPs see \cite{Wiersig:2025woc}), only one eigenvector $r_0$ remains. Their two-dimensional space is completed by a second vector $r_1$ satisfying
\begin{equation}
 (L_N-\omega_{\rm EP}I)r_0=0,\qquad
 (L_N-\omega_{\rm EP}I)r_1=r_0,
 \label{eq:Aminus2-chain}
\end{equation}
where the two vectors form a Jordan chain. We denote by $\Pi_{\rm EP}$ the spectral projection onto their two-dimensional space, excluding all other modes. This projection need not be orthogonal with respect to the energy inner product. The matrix
\begin{align}
 A_{-2}&=(L_N-\omega_{\rm EP}I)\Pi_{\rm EP},\quad
 A_{-2}\ne0,\quad A_{-2}^2=0,
 \label{eq:Aminus2-nilpotent}\\
 A_{-2}r_0&=0,\qquad A_{-2}r_1=r_0
 \label{eq:Aminus2-action}
\end{align}
describes the coupling between the two EP states. The full local expansion of the resolvent is then
\begin{equation}
 R_N(z)=\frac{A_{-2}}{(z-\omega_{\rm EP})^2}
       +\frac{\Pi_{\rm EP}}{z-\omega_{\rm EP}}
       +R_{\rm reg}(z),
 \label{eq:Aminus2-Laurent}
\end{equation}
where $R_{\rm reg}$ is analytic near the pair. The double pole gives
\begin{align}
 \|R_N(z)\|_E&\sim
 \frac{\|A_{-2}\|_E}{|z-\omega_{\rm EP}|^2},
 \label{eq:Aminus2-resolvent-strength}\\
 |z-\omega_{\rm EP}|&\sim
 \sqrt{\epsilon\,\|A_{-2}\|_E},\qquad \epsilon\to0^+.
 \label{eq:Aminus2-pseudospectral-radius}
\end{align}
The second relation gives the local contour radius at the exact EP of this matrix~\cite{Cao2026}. The double pole fixes the square-root exponent, while the prefactor depends on the energy norm and includes the spectral projection.

Along a generic parameter path approaching an EP at fixed discretization, the pair separation and condition numbers scale as
\begin{equation}
 d(t)\propto|t|^{1/2},\qquad
 \kappa_{1,2}(t)\propto|t|^{-1/2}.
 \label{eq:paper-ep-scaling}
\end{equation}
Here $t$ is a background-parameter offset, not physical time. For a local frequency condition $F(z,t)=0$ with $F=F_z=0$, $F_{zz}\ne0$ and $F_t\ne0$ at the EP,
\begin{equation}
 \omega_\pm(t)=\omega_{\rm EP}
       \pm\sqrt{-\frac{2F_t}{F_{zz}}t}+\mathcal O(t).
 \label{eq:generic-unfolding}
\end{equation}
All derivatives are evaluated at the EP. A path with $F_t=0$ can instead produce a different scaling. As the modes coalesce, their simple-mode condition numbers diverge. A finite value of this condition number cannot be assigned to the exact double root.

For two distinct modes close to the EP, the collective resolvent can be written exactly by setting $\omega_c=(\omega_++\omega_-)/2$, $\delta=(\omega_+-\omega_-)/2$, and $J=(L_N-\omega_c I)\Pi$, where $\Pi$ selects the isolated pair:
\begin{align}
 J^2&=\delta^2\Pi,\qquad
 R_N(z)\Pi=\frac{(z-\omega_c)\Pi+J}{(z-\omega_c)^2-\delta^2},
 \label{eq:pair-resolvent}\\
 \Pi_\pm&=\tfrac12(\Pi\pm J/\delta).
 \label{eq:pair-projectors}
\end{align}
Although the individual projections can grow as $1/|\delta|$, the combined response approaches a finite double-pole limit at fixed $z\ne\omega_{\rm EP}$. The separate simple-pole description applies when $\kappa_j\epsilon\ll d$. At larger thresholds, merging contours describe the response of the pair as a whole, and their areas need not follow the ordering of the individual condition numbers. For the scalar branch studied below, the angular functions depend on frequency. The calculation must account for that dependence in the state and residual maps, as a single fixed angular matrix does not describe it.

\section{Metric, scalar energy and angular-branch restriction}
\label{sec:scalar}
\subsection{Kerr--Newman metric}

We consider a subextremal KN BH of mass $M$, electric charge $Q$ and angular momentum $J=aM$, with $M^2>a^2+Q^2$. In Boyer--Lindquist coordinates $(t,r,\theta,\phi)$, the metric $g$, with signature $(-,+,+,+)$, has a line element
\begin{align}
 ds^2=g_{ab}\,dx^a dx^b
 ={}&-\left(1-\frac{2Mr-Q^2}{\Sigma}\right)dt^2
 \nonumber\\
 &-\frac{2a(2Mr-Q^2)\sin^2\theta}{\Sigma}\,dt\,d\phi
 \nonumber\\
 &+\frac{\Sigma}{\Delta}\,dr^2+\Sigma\,d\theta^2
 +\frac{\mathcal A\sin^2\theta}{\Sigma}\,d\phi^2,
 \label{eq:knx-metric}
\end{align}
where
\begin{align}
 \Sigma&=r^2+a^2\cos^2\theta,\nonumber\\
 \Delta&=r^2-2Mr+a^2+Q^2,\nonumber\\
 \mathcal A&=(r^2+a^2)^2-a^2\Delta\sin^2\theta.
 \label{eq:knx-functions}
\end{align}
The event horizon radius lies at $r=r_+=M+\sqrt{M^2-a^2-Q^2}$, while the Cauchy horizon radius lies at $r=r_-=M-\sqrt{M^2-a^2-Q^2}$, with exterior region $r>r_+$. The mass $M$ has been restored in these background expressions. We set $M=1$ again below.

\subsection{Scalar energy on hyperboloidal slices}

The neutral massless field obeys $\Box_g\Phi=0$. We set $R=r_+$ and compactify the exterior with $x=R/r$. The future event horizon is fixed at $x=1$ and future null infinity is at $x=0$. The Cauchy horizon remains outside this computational domain. The hyperboloidal coordinates and regular field are
\begin{align}
 T&=t+r_*+4\ln x-2r,\qquad \varphi=\phi+k(r),\nonumber\\
 r_*'&=\frac{r^2+a^2}{\Delta},\qquad k'=\frac a\Delta,
 \qquad \Phi=\frac{e^{im\varphi}}{r\sqrt{2\pi}}u.
 \label{eq:scalar-coordinates}
\end{align}
The slices cross the future horizon and extend to future null infinity. Their geometry and endpoint limits are derived in Appendix~\ref{app:hypkn}. The transformed wave equation and its first-order evolution form are
\begin{align}
 P(z)&=P_0+zP_1+z^2P_2,\nonumber\\
 L&=\begin{pmatrix}0&I\\-P_2^{-1}P_0&-P_2^{-1}P_1\end{pmatrix},
 \nonumber\\
 y&=\binom{u}{i\partial_Tu},\qquad \partial_Ty=-iLy.
 \label{eq:scalar-companion}
\end{align}
Appendix~\ref{app:scalar-validation} lists the coefficients used in the radial discretization, and Appendix~\ref{sec:domain} discusses the outgoing-domain requirements. Since the frequency is conjugate to $T$, restoring dimensions gives $z=M\omega$, with no additional rescaling by the horizon radius.

The scalar stress-energy tensor defines the energy flux through a hyperboloidal slice $\Sigma_T$ with future unit normal $n^a$:
\begin{align}
 T_{ab}&=\Rew(\nabla_a\bar\Phi\nabla_b\Phi)
       -\tfrac12g_{ab}\nabla_c\bar\Phi\nabla^c\Phi,\nonumber\\
 E_X[\Phi]&=\int_{\Sigma_T}T_{ab}X^an^b\,\dd\Sigma_T.
 \label{eq:energy}
\end{align}
We choose the future-directed vector field
\begin{align}
 X&=\partial_t+\Omega_Z\partial_\phi,\nonumber\\
 \Omega_Z&=-\frac{g_{t\phi}}{g_{\phi\phi}}
          =\frac{a(2r-Q^2)}{\mathcal A},
 \label{eq:multiplier}\\
 g(X,X)&=-\frac{\Delta\Sigma}{\mathcal A}.
 \label{eq:knx-norm}
\end{align}
The condition $g(X,\partial_\phi)=0$ identifies the vector with the direction of locally nonrotating observers of zero angular momentum. It is timelike throughout the open exterior, tends to $\partial_t$ at infinity and approaches the null horizon generator $\partial_t+\Omega_H\partial_\phi$ at $r=r_+$, where $\Omega_H=a/(r_+^2+a^2)$.

By the dominant energy condition, the flux is nonnegative. The observers' unit velocity, $u_Z=X/\sqrt{-g(X,X)}$, is normal to Boyer--Lindquist constant-$t$ slices and differs from the unit normal $n^a$ of the hyperboloidal slices. In a rotating spacetime, $\Omega_Z$ varies with position and $X$ is generally not a Killing vector. The resulting energy has an inner product term in its balance law. It, nevertheless, provides a positive measure in the presence of the ergoregion~\cite{Cai2025} and reduces to the positive stationary scalar energy at $a=0$. The weight degenerates at the horizon, but the norm is positive definite on the finite trial space for nonzero $m$. We denote its discrete matrix by $G_N$. Positivity is established in Appendix~\ref{app:hypkn}, and the energy density and quadrature are given in Appendix~\ref{app:scalar-validation}.

\subsection{The frequency-dependent \texorpdfstring{$\ell=m=1$}{l=m=1} sector}
\label{sec:branch}

For $a\ne0$, the spheroidal function $S_{11}(\theta;az)$ contains contributions from the spherical harmonics $\ell=1,3,5,\ldots$. Selecting one spherical harmonic would not isolate the separated branch. We first discretize the radial coordinate in Eq.~\eqref{eq:scalar-companion} without truncating the angular functions, and denote the resulting operator by $L_{N,m}$. Its factor $P_2^{-1}$ acts at the radial nodes before radial interpolation.

At a fixed test frequency $z$, the trial state contains independent field and velocity data,
\begin{align}
 q&=\binom{\boldsymbol u}{\boldsymbol v},\qquad v=i\partial_Tu,\nonumber\\
 \boldsymbol u&=(u(x_0),\ldots,u(x_{N-1}))^T,\nonumber\\
 \boldsymbol v&=(v(x_0),\ldots,v(x_{N-1}))^T.
 \label{eq:trial-state-coordinates}
\end{align}
The vector $q\in\mathbb C^{2N}$ contains spatial field and velocity data. It is distinct from $z$, the frequency at which these data are tested. A QNM satisfies $v=zu$, whereas a general trial state need not satisfy this relation. The map
\begin{equation}
 J_1(z)q=
 \begin{pmatrix}
 u(x)S_{11}(\theta;az)\\
 v(x)S_{11}(\theta;az)
 \end{pmatrix}
 \label{eq:trial-state-injection}
\end{equation}
attaches the chosen angular function to both radial interpolants. The subscript $1$ identifies the $\ell=1$ branch with $m=1$ fixed. The minimum normalized residual on this branch is
\begin{equation}
 \sigma_{1,E_X}(z)=\min_{q\ne0}
 \frac{\|(zI-L_{N,m})J_1(z)q\|_{E_X}}
      {\|J_1(z)q\|_{E_X}}.
 \label{eq:scalar-parent-residual}
\end{equation}
Equivalently, one may minimize over states normalized by $\|J_1(z)q\|_{E_X}=1$. The minimizing configuration comes closest to satisfying the QNM equation at the test frequency. The definition applies to any background, including the baseline and near-EP cases considered here. Where the unrestricted resolvent exists, the reciprocal of this residual measures the gain for sources in $(zI-L_{N,m})\operatorname{range}J_1(z)$. Both the angular subspace and the allowed sources depend on frequency. When $a=0$, the angular subspace is invariant under the evolution operator.

The plotted finite residual uses a continued-fraction angular eigenvalue and numerical angular integrals of the energy. In coordinates $(u,f)$ with $f=zu-v$, it is
\begin{align}
 D_1(z)&=\begin{pmatrix}0&I\\P_{\rm br}(z)&-P_{1,r}\end{pmatrix},\nonumber\\
 \sigma_1^{G_N}(z)&=s_{\min}
 \!\left[C_r(z)D_1(z)C_y(z)^{-1}\right].
 \label{eq:constrained}
\end{align}
Here $P_{\rm br}(z)$ is the separated radial operator along the continued $\ell=1$ angular branch. The matrix $P_{1,r}$ is the coefficient of the linear frequency term before the angular dependence is eliminated. The factors $C_y$ and $C_r$ express the energies of the reconstructed state and residual as Euclidean norms. These factors generally differ, since the state and residual are reconstructed by different maps. The maps and their finite angular approximations are detailed in Appendix~\ref{app:scalar-validation}.

We refer to the regions where $\sigma_1^{G_N}(z)<\epsilon$ as the \emph{branch-constrained energy pseudospectrum}. Equation~\eqref{eq:constrained} is a finite approximation to Eq.~\eqref{eq:scalar-parent-residual}. Because the angular function changes with frequency, this construction cannot be represented by one frequency-independent fixed-$\ell$ evolution matrix. We continue the positive- and negative-real-frequency roots independently through the mode interactions~\cite{Davey2023}.

\section{Scalar pseudospectra and exceptional-point condition number}
\label{sec:results}

\begin{figure*}[t]
 \centering
 \includegraphics[width=0.99\textwidth]{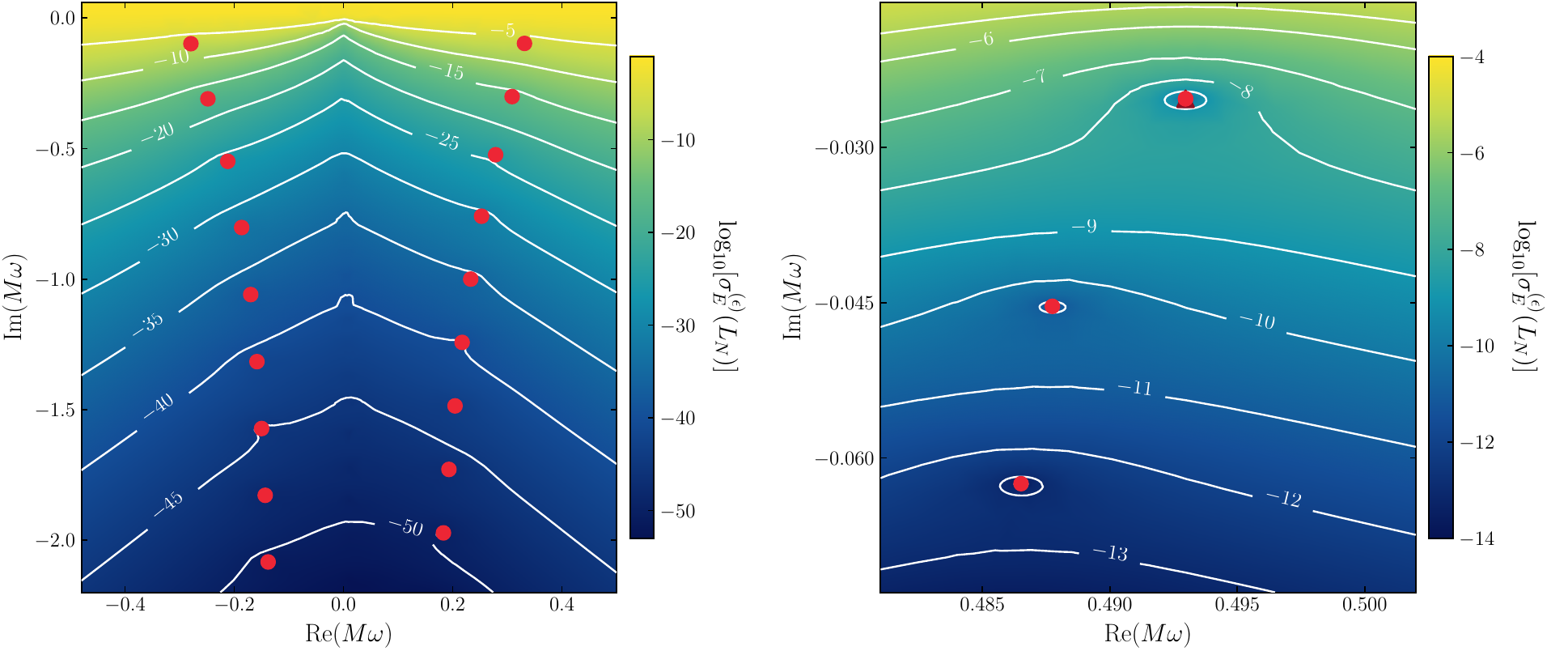}
\caption{Branch-constrained scalar $\ell=m=1$ energy-norm pseudospectra $\sigma_E^{(\epsilon)}(L_N)$. The log-scale colorbar corresponds to various choices of $\epsilon$, which are shown in the contour plot as $\mathrm{log}_{10}(\epsilon)$. Left: $(a/M,Q/M)=(0.3,0.4)$, with the independently continued positive- and negative-real-frequency QNMs through $n=8$. Right: The near-extremal background Eq.~\eqref{eq:scalar-enhanced-background}, with the close pair and two more-damped QNMs. Filled red dots mark QNMs. The pair overlaps visually in the wide view. The dark red triangle denotes the finite-grid EP at its different critical background, not a further eigenvalue of the plotted operator. Both maps use $N=200$ and the physical energy in Eqs.~\eqref{eq:energy} and \eqref{eq:constrained}. Numerical settings and interpolation checks are given in Appendix~\ref{app:scalar-validation}. The central cusps belong to the unmarked finite residual structure, i.e., no continuum branch-cut identification is inferred from their shape.}
 \label{fig:scalarbaseline}
 \label{fig:scalarepwide}
\end{figure*}

All pseudospectra shown here use $N=200$ radial collocation points. The scalar colorbars use $\sigma_E^{(\epsilon)}(L_N)$ defined in Eq.~\eqref{eq:constrained}. Figure~\ref{fig:scalarbaseline} (left) shows the $\ell=m=1$ scalar pseudospectrum for a KN BH with $(a/M,Q/M)=(0.3,0.4)$. The broad contours indicate spectral instability, with frame dragging producing an asymmetry between the positive- and negative-real-frequency branches. Although the nonconvergent discretized modes at the imaginary axis are not marked, their residual structure remains visible in the contours. The right panel shows the weakly damped modes of a near-extremal KN BH. The two upper red markers form the near-EP pair and overlap at this scale. The targeted EP joins the fundamental and first-overtone branches of the $(\ell,m)=(1,1)$ sector~\cite{CavalcanteRichartz2026}. The EP lies at
\begin{equation}
 \begin{gathered}
 g_{\rm EP}^{(N)}\equiv(a/M,Q/M)_{\rm EP}^{(N)},\\
g_{\rm EP}^{(N)}\simeq(\ScalarEPa,\ScalarEPQ),\\
 M\omega_{\rm EP}^{(N)}\simeq
       \ScalarEPOmegaRe-\ScalarEPOmegaIm i.
 \end{gathered}
 \label{eq:scalar-ep-reference}
\end{equation}
The plotted operator is evaluated at a nearby background rather than at the EP itself. We choose
\begin{equation}
 \begin{gathered}
 g(t)=g_{\rm EP}^{(N)}+t\bigl[g_{\rm ref}-g_{\rm EP}^{(N)}\bigr],\\
 g_{\rm ref}=(0.791540,0.610450),\qquad t=\ScalarApproachT.
 \end{gathered}
 \label{eq:scalar-enhanced-background}
\end{equation}
The coefficients are evaluated using unrounded parameter values. At this background, the two roots, labelled near-EP 1 and near-EP 2, have the nonzero separation $d=|M(\omega_1-\omega_2)|=\ScalarPairGap$.

\begin{figure}[t]
 \centering
 \includegraphics[width=\columnwidth]{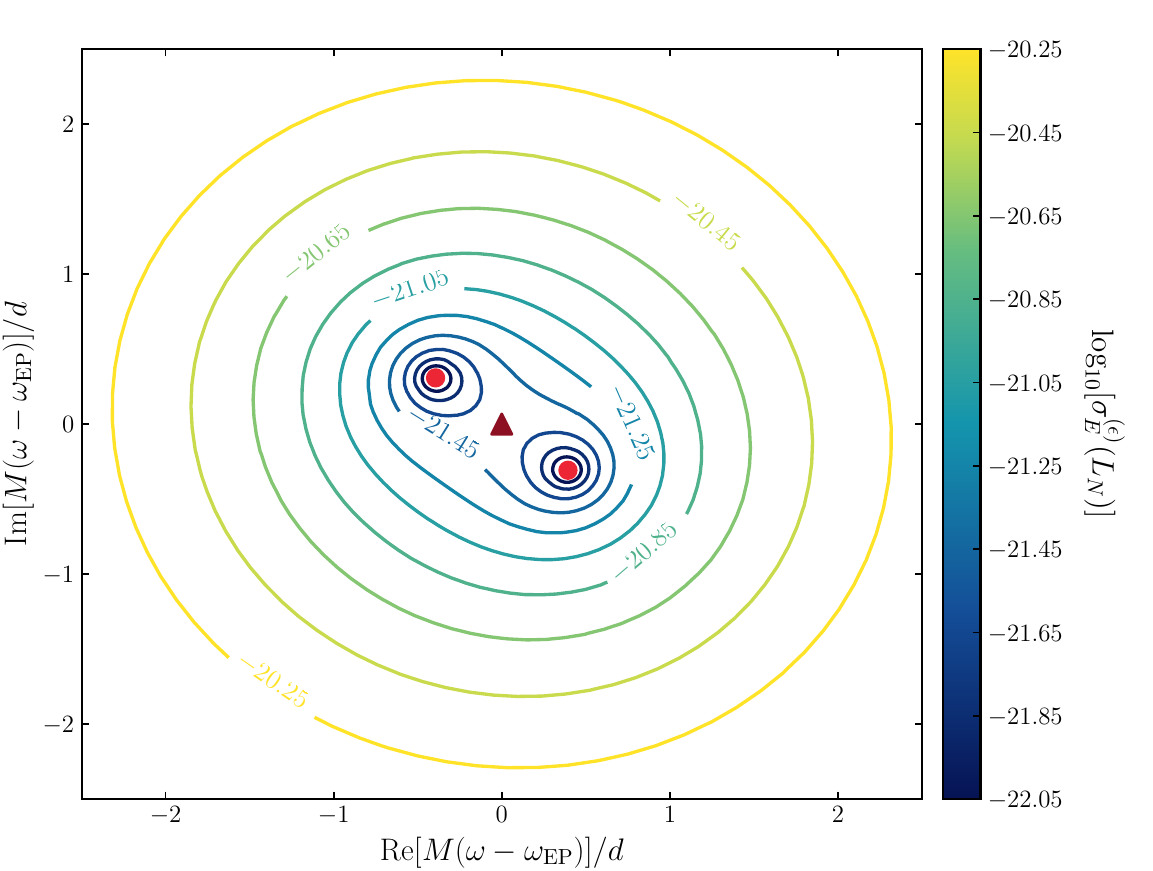}
\caption{Local scalar view at the same background and resolution as Fig.~\ref{fig:scalarbaseline} (right). Coordinates are $\Rew[M(\omega-\omega_{\rm EP}^{(N)})]/d$ and $\Imw[M(\omega-\omega_{\rm EP}^{(N)})]/d$, where $d=\ScalarPairGap$. This magnifies frequency coordinates only: singular residuals and $\epsilon$ contours are not rescaled. Ten colored contours, with their own threshold range, resolve two distinct roots designated by red dots. The triangle is the critical-background reference. Contour labels give $\log_{10}\sigma_E^{(\epsilon)}(L_N)$. The colorbar uses the same notation as Fig.~\ref{fig:scalarbaseline}.}
 \label{fig:scalareplocal}
\end{figure}

The local view in Fig.~\ref{fig:scalareplocal} resolves two response regimes. At the smallest thresholds, each simple pole is surrounded by separate contours. These merge at larger thresholds, where the response becomes collective~\cite{Cao2026}. The contour merger occurs while the near-EP operator still has two distinct eigenvalues.

For the selected scalar pair, the simple-pole validity criterion gives
\begin{equation}
 \epsilon_{\mathrm{cross},\mathrm{sc}}
 \sim\frac{d}{\kappa_{1,E_X}}
 \simeq1.06\times10^{-21}.
\end{equation}
This scale falls within the threshold range of Fig.~\ref{fig:scalareplocal}, in agreement with the transition from separate to collective contours. It estimates the crossover only to order of magnitude and does not fix the precise merger threshold.

\subsection{Condition number and spectral instability}
\label{sec:conditioning}

\begin{figure*}[t]
 \centering
 \includegraphics[width=0.98\textwidth]{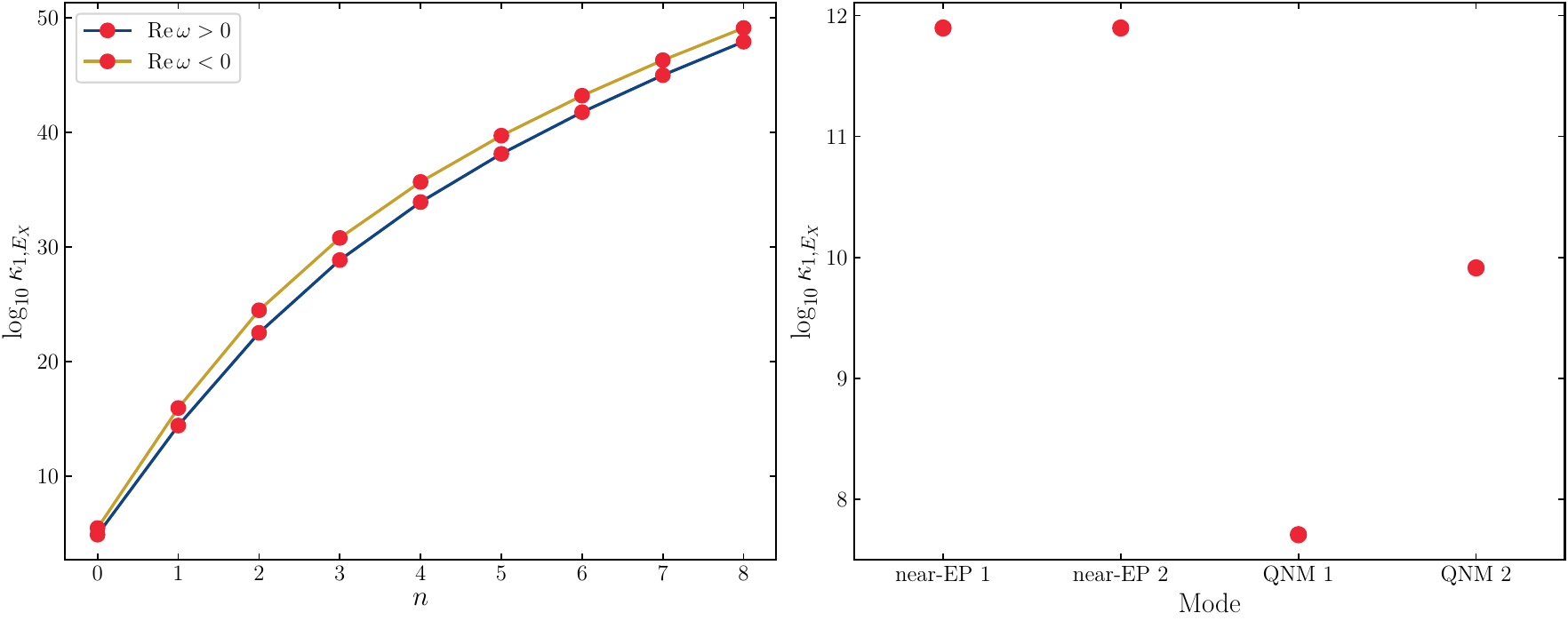}
\caption{Scalar branch, energy-norm, condition numbers from Eq.~\eqref{eq:branchkappa}, with $N=200$ and $\ell=m=1$. Left: the scalar QNM positive- and negative-real-frequency branches through $n=8$. Right: near-EP 1, near-EP 2 and the two more-damped QNMs at Eq.~\eqref{eq:scalar-enhanced-background}. At this offset and resolution, both near-EP coefficients exceed those of the two displayed comparison modes. Figure~\ref{fig:resolution-ordering} shows how this ordering depends on $N$.}
 \label{fig:conditioning}
\end{figure*}

We evaluate the local pole coefficient with the state and residual energy matrices used in Eq.~\eqref{eq:constrained}. Writing $P(z)=P_{\rm br}(z)$, let $P(\omega_j)r_j=0$ and $l_j^\dagger P(\omega_j)=0$. In terms of the coordinates and energy factors defined in Eqs.~\eqref{eq:scalar-pullback-maps}--\eqref{eq:scalar-pullback-grams}, the energy-norm condition number is
\begin{equation}
 \kappa_{1,E_X}(\omega_j)=
 \frac{\left\|C_y\binom{r_j}{0}\right\|_2
       \left\|C_r^{-\dagger}\binom{P_{1,r}^\dagger l_j}{l_j}\right\|_2}
      {|l_j^\dagger P'(\omega_j)r_j|}.
 \label{eq:branchkappa}
\end{equation}
The energy factors are evaluated at the pole, and the total derivative in Eq.~\eqref{eq:scalar-branch-pencil} includes the dependence of the angular eigenvalue on frequency. Equation~\eqref{eq:branchkappa} is invariant under rescaling either mode vector and incorporates both the state and residual energies~\cite{Tisseur2000}. Near a simple mode,
\begin{equation}
 \sigma_1^{G_N}(z)=\frac{|z-\omega_j|}{\kappa_{1,E_X}(\omega_j)}
                  +\textrm{regular terms}.
 \label{eq:branchslope}
\end{equation}
The coefficient measures the local response on the selected angular branch. The unrestricted response of a fixed angular evolution matrix is a different quantity. The inverse-map derivation and sampled slope checks appear in Appendix~\ref{app:scalar-validation}.

At the moderate background, the positive- and negative-real-frequency fundamental modes ($n=0$) have coefficients $8.13\times10^4$ and $3.07\times10^5$, respectively. The coefficients grow along both branches, reaching $8.05\times10^{47}$ and $1.26\times10^{49}$ at $n=8$, as shown in Fig.~\ref{fig:conditioning} and Table~\ref{tab:branchconditions}. Both near-EP coefficients are approximately $7.90\times10^{11}$, whereas the two displayed, more-damped modes have values $5.07\times10^7$ and $8.18\times10^9$. The smaller pair coefficient exceeds the larger comparison value by a factor of about $96.6$, with the background, norm and radial resolution held fixed. This factor depends on the chosen offset and resolution, and it decreases as $N$ increases (Sec.~\ref{sec:resolution-tests}).

These coefficients depend strongly on radial resolution even when the QNM frequencies have converged accurately. For the positive baseline fundamental QNM, $N=160,200,240$ give approximately $1.77\times10^4$, $8.13\times10^4$, $3.36\times10^5$. The reported condition numbers characterize the finite discretized problem, for which convergence of the frequencies alone is insufficient.

\begin{table*}[t]
\centering\small
\begin{minipage}[t]{0.33\textwidth}
\centering
\begin{tabular}[t]{rcc}
\hline
$n$ & $\kappa_{1,E_X}(\omega_+)$ & $\kappa_{1,E_X}(\omega_-)$ \\
\hline
0 & $8.1276\times10^{4}$ & $3.0726\times10^{5}$ \\
1 & $2.6768\times10^{14}$ & $9.0887\times10^{15}$ \\
2 & $3.2921\times10^{22}$ & $3.0420\times10^{24}$ \\
3 & $7.2941\times10^{28}$ & $6.1651\times10^{30}$ \\
4 & $8.2997\times10^{33}$ & $4.7467\times10^{35}$ \\
5 & $1.3397\times10^{38}$ & $5.1465\times10^{39}$ \\
6 & $5.7763\times10^{41}$ & $1.5702\times10^{43}$ \\
7 & $9.7104\times10^{44}$ & $1.9678\times10^{46}$ \\
8 & $8.0524\times10^{47}$ & $1.2620\times10^{49}$ \\
\hline
\end{tabular}
\end{minipage}\hfill
\begin{minipage}[t]{0.65\textwidth}
\centering
\begin{tabular}[t]{l@{\quad}c@{\quad}c}
\hline
Mode & $M\omega$ & $\kappa_{1,E_X}$ \\
\hline
near-EP 1 & $0.492968350301-0.025307752950i$ & $7.8980745\times10^{11}$ \\
near-EP 2 & $0.492968350960-0.025307753465i$ & $7.8980761\times10^{11}$ \\
QNM 1 & $0.487742751102-0.045361184615i$ & $5.0749316\times10^{7}$ \\
QNM 2 & $0.486514258755-0.062498988873i$ & $8.1789301\times10^{9}$ \\
\hline
\end{tabular}
\end{minipage}
\caption{All $22$ selected scalar branch condition numbers from Eq.~\eqref{eq:branchkappa}, at $N=200$. Left: the positive- and negative-real-frequency branches at $(a,Q)=(0.3,0.4)$, through $n=8$. Right: four $\ell=m=1$ QNMs and corresponding condition numbers at Eq.~\eqref{eq:scalar-enhanced-background}. The local labels near-EP 1 and near-EP 2 refer to the fundamental/first-overtone branches. QNM 1 and QNM 2 denote only the two displayed more-damped modes.}
\label{tab:branchconditions}
\end{table*}

\section{Coupled gravitoelectromagnetic perturbations}
\label{sec:ge}

\subsection{A fixed radiative state and a positive energy}

Although the separated equations \cite{Hintz2026,Destounis2026} reproduce the coupled-system QNMs \cite{Pombo:2026xsj,DiasEtAl2022}, an energy norm requires more than the QNM frequencies. We work with the positive-spin four-field system of Ref.~\cite{Hintz2026} before angular separation. Appendix~\ref{app:ge-energy} defines its curvature amplitudes $\psi_2,\psi_1$ and connection amplitudes $k_{\rm NP},s_{\rm NP}$ in a tetrad for which the coupled fields include electromagnetic radiation. We choose the relative normalization
\begin{equation}
 U=(\psi_2,\psi_1,2Qk_{\rm NP},2Qs_{\rm NP})^T.
 \label{eq:ge-base-state}
\end{equation}
The factor $2Q$ fixes the field normalization used in both the energy and the source conversion. All backgrounds considered here have $Q\ne0$.

Using the same hyperboloidal coordinates as for the scalar field, with $R=r_+$, $c_0=a^2+Q^2$ and $x=R/r$, the regular four-field state is
\begin{equation}\label{eq:ge-regular-state}
 Y=r(\Delta^2\psi_2,\Delta\psi_1,
                   2Q\Delta^2k_{\rm NP},2Q\Delta s_{\rm NP})^T.
\end{equation}
The transformed equations take the form
\begin{equation}
\label{eq:ge-hyp-system}
 B\partial_TY+\mathsf A^x\partial_xY+
 \mathsf A^\theta\partial_\theta Y+
 \mathsf A^\varphi\partial_\varphi Y+\mathsf C Y=0.
\end{equation}
The time coefficient consists of two Hermitian blocks,
\begin{equation}\label{eq:ge-Bblock}
 \begin{gathered}
 B_\eta=\begin{pmatrix}
 b&-i\,\eta\, a\sin\theta\\ i\,\eta\, a\sin\theta&d
 \end{pmatrix},\qquad \eta=\pm1,\\
 b=2+\frac4r,\qquad
 d=8-2Q^2-\frac{4c_0}{r}.
 \end{gathered}
\end{equation}
Here $\eta=+1$ labels the Weyl block and $\eta=-1$ the connection block. Since $c_0/r\le r_-<1$, both blocks satisfy
\begin{equation}
\label{eq:ge-uniform-bound}
 1<2-|a|\le\lambda_{\min}(B_\eta)
 \le\lambda_{\max}(B_\eta)<9,
\end{equation}
where $\lambda_{\rm min,max}$ are the minimum and maximum eigenvalues of $B_\eta$. Positivity holds throughout the compactified domain, including the horizon and null infinity, and gives the frequency-independent radiative energy
\begin{equation}\label{eq:ge-energy}
 \|Y\|_E^2=\frac12\int_0^1\!\dd x\int_0^\pi\!\sin\theta\,\dd\theta
             \int_0^{2\pi}\!\dd\varphi\;Y^\dagger B Y.
\end{equation}
The balance law contains an outgoing semidefinite boundary flux and an inner product term whose magnitude is bounded by a constant times the instantaneous energy. Appendix~\ref{app:ge-energy} gives the derivation and the regular coefficient matrices. This energy measures the chosen radiative fields. Its relation to metric--Maxwell energy and physical sources is discussed in Sec.~\ref{sec:discussion}.

\subsection{The generator and its forcing norm}

In the $e^{im\varphi}$ sector, we define the operators
\begin{equation}\label{eq:ge-generator}
 \begin{gathered}
 \mathcal D_m=\mathsf A^x\partial_x+\mathsf A^\theta\partial_\theta
                         +im\mathsf A^\varphi+\mathsf C,\\
 \mathcal L_m=-iB^{-1}\mathcal D_m,\qquad
 \partial_TY=-i\mathcal L_mY.
 \end{gathered}
\end{equation}
Integration of Eq.~\eqref{eq:ge-energy} on the numerical basis gives $G_N$. Applying the definitions of Sec.~\ref{sec:framework}, we use $\sigma_E(z)$ for the resulting minimum residual. This quantity measures the full fixed-$m$ response across all angular families included in the calculation. The scalar maps describe the selected spheroidal branch.

Each radial amplitude is represented on $N=200$ collocation points. The angular expansion uses spin-weighted spherical Jacobi bases with spins $(2,1,1,2)$ and maximum angular harmonic degree $L_{\rm ang}=6$, as detailed in Appendix~\ref{app:numerics}. The resulting discretized operator has dimension $4400$ for $m=1$ and $4000$ for $m=2$. Its discrete form is
\begin{equation}
\label{eq:ge-numerical-pencil}
 P_N(z)=D_N-izB_N,\qquad L_N=-iB_N^{-1}D_N,
\end{equation}
so the generator resolvent is
\begin{equation}\label{eq:ge-forced-inverse}
 (zI-L_N)^{-1}=-iP_N(z)^{-1}B_N,
\end{equation}
Here $D_N$ is the discretization of $\mathcal D_m$ in Eq.~\eqref{eq:ge-generator}. A state-rate source $f$ means $B_N\dot Y+D_NY=B_Nf$.
For a real driving frequency $\Omega$ outside the finite spectrum, the monochromatic particular response is
\begin{equation}\label{eq:ge-harmonic-forcing}
 \dot Y=-iL_NY+fe^{-i\Omega T},\qquad
 \widehat Y=i(\Omega I-L_N)^{-1}f.
\end{equation}
At real frequencies, the resolvent measures the largest state response to this forcing. The complex-frequency maps extend the same response measure away from the real axis.

The matrix $B_N$ converts state forcing into a source of the stored equations and includes the equation and basis transformations applied to $P_N$. The same conversion enters the condition number. If $w_j$ is a left vector of the stored generalized eigenvalue problem, the left vector of $L_N$ is $y_j=B_N^\dagger w_j$. The numerical formula is given in Appendix~\ref{app:numerics}.

The roots, left and right vectors, and energy factors are computed with $260$ working digits. Arithmetic precision, stopping tolerance and truncation accuracy are controlled separately. Appendix~\ref{app:numerics} summarizes the independent dense, limiting-frequency and direct-resolvent checks.

\section{Gravitoelectromagnetic pseudospectra near a coupled exceptional point}
\label{sec:maps}

We follow the GE EP reported in Ref.~\cite{MisasArcos2026}. With $g=(a/M,Q/M)$, we denote by $g_{\rm EP}^{(N)}$ the double-root candidate of the generalized eigenvalue problem at fixed $N=200$ and $L_\textrm{ang}=6$. We approach the candidate along the path
\begin{equation}\label{eq:ge-off-ep}
 \begin{aligned}
 a_{\rm ref}&=0.3041149160474578,\\
 Q_{\rm ref}&=0.9523576757342892,\\
 g_{\rm ref}&=(a_{\rm ref},Q_{\rm ref}),\\
 g(t)&=g_{\rm EP}^{(N)}+t[g_{\rm ref}-g_{\rm EP}^{(N)}],\quad m=2.
 \end{aligned}
\end{equation}
The candidate frequency is
\begin{equation}\label{eq:ge-ep-reference}
 \begin{aligned}
 M\omega_{\rm EP}^{(N)}&\simeq\GEepOmegaRe\\
 &\quad-\GEepOmegaIm i,
 \end{aligned}
\end{equation}
at spin $a/M\simeq\GEepA$ and charge $Q/M\simeq\GEepQ$. The calculation uses unrounded values throughout. The checks in Appendix~\ref{app:numerics} support the coalescence of two frequencies with a single independent eigenvector in this discretized problem.

\begin{figure*}[t]
 \centering
 \includegraphics[width=0.99\textwidth]{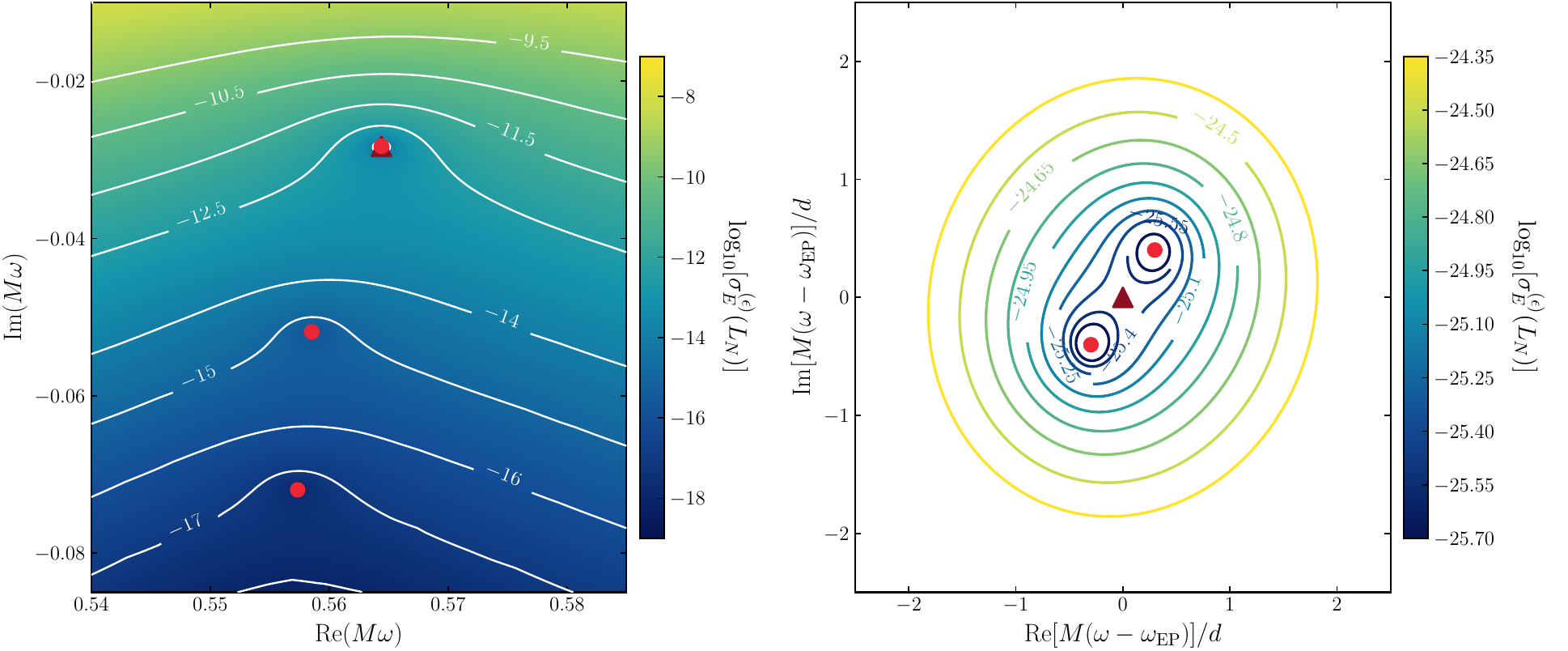}
\caption{Coupled GE energy-norm pseudospectra $\sigma_E^{(\epsilon)}(L_N)$, Eq.~\eqref{eq:ge-off-ep}. The log-scale colorbar corresponds to various choices of $\epsilon$, which are shown in the contour plot as $\mathrm{log}_{10}(\epsilon)$ at $g(\GEapproachT)$. Here, $m=2$, $N=200$ and $L_{\rm ang}=6$, in the radiative energy \eqref{eq:ge-energy}. Left: spectral neighborhood of the near-EP pair and two subdominant QNMs, using $\GEwideSamples$ measured frequencies. The pair markers overlap at this scale. Ten white contours overlay the $\log_{10}\sigma_E^{(\epsilon)}(L_N)$ field. Right: the same pair in coordinates $\Rew[M(\omega-\omega_{\rm EP}^{(N)})]/d$ and $\Imw[M(\omega-\omega_{\rm EP}^{(N)})]/d$, where $d=\GEpairGap$, using $\GElocalSamples$ measured frequencies. Ten colored contours span $-25.70\le\log_{10}\epsilon\le-24.35$ without a filled plot colormap. Filled red dots mark QNMs, while the dark red triangle denotes the critical EP reference \eqref{eq:ge-ep-reference} at its different background. Pole-factored interpolation uses the measured full resolvent, without a modal truncation (see Appendix~\ref{app:numerics}).}
 \label{fig:ge-pair}
 \label{fig:ge-local}
\end{figure*}

Figure~\ref{fig:ge-pair} shows the full coupled pseudospectrum at $t=\GEapproachT$. The two simple eigenvalues labeled near-EP 1 and near-EP 2 have separation $d=|M(\omega_1-\omega_2)|=\GEpairGap$. QNM 1 and QNM 2 are two subdominant eigenvalues at the same background. These local labels do not assign global overtone indices. The left panel places the close pair in its spectral neighborhood, and the magnified right panel resolves the splitting. Division by $d$ rescales the frequency coordinates only. The singular residual and $\sigma_E^{(\epsilon)}(L_N)$ keep their original normalization.

In the right panel, the two lowest contour levels surround the near-EP QNMs separately, and the eight higher levels enclose them jointly. The transition reflects the collective response of the close pair. As in the scalar case, merging contours at finite $\epsilon$ do not establish an EP. The plotted operator has distinct eigenvalues, and the triangle marks the critical reference at different KN parameters. The interpolation uses analytic pole factors to regularize the measured full resolvent, without replacing it by a sum of four pole residues. The pair gap and condition numbers give the characteristic crossover scale
\begin{equation}\label{eq:ge-crossover-scale}
 \epsilon_{\rm cross}\sim \frac{d}{\kappa_E}\simeq1.33\times10^{-25}.
\end{equation}
This estimate lies within the local contour range, explaining why the separate-pole regime appears only at very small thresholds. The precise merger threshold also depends on the full two-mode contribution and the regular response of the remaining modes.

\subsection{Condition numbers and a matched approach to coalescence}
\label{sec:measured-conditioning}

\begin{figure*}[t]
 \centering
 \includegraphics[width=0.99\textwidth]{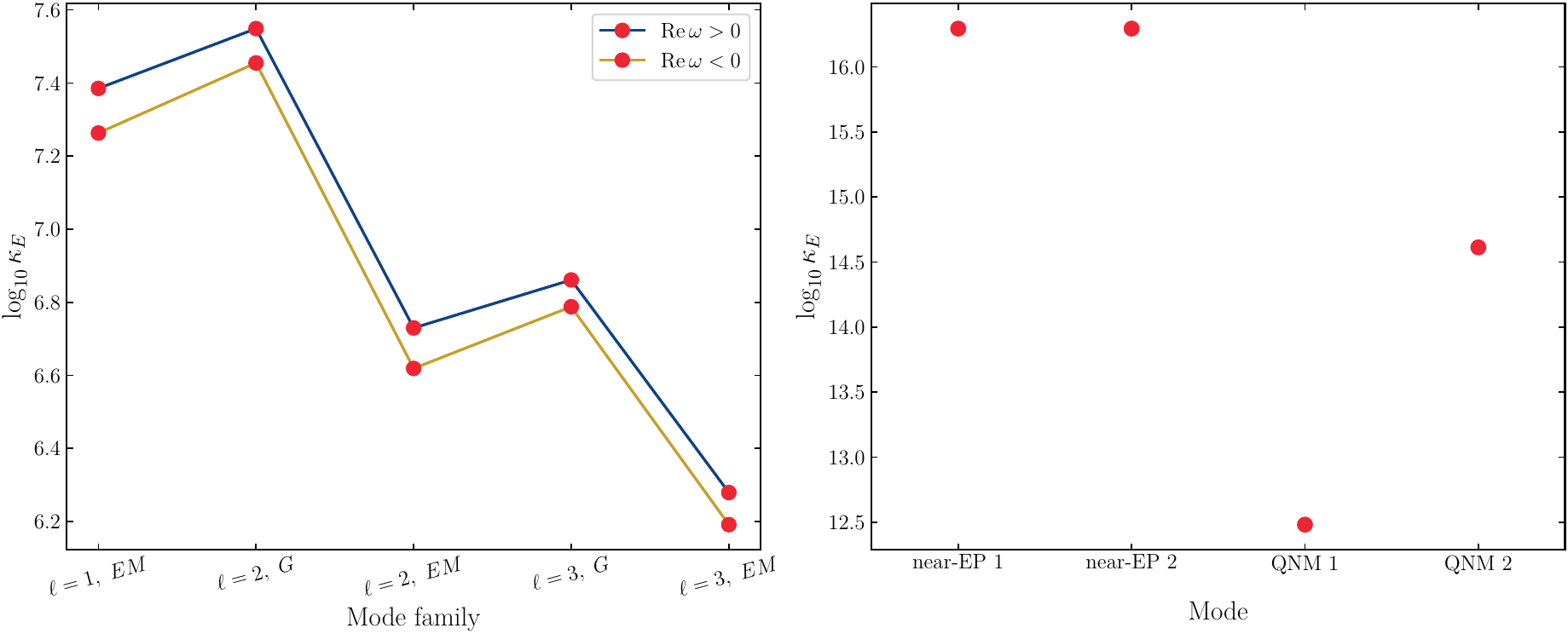}
\caption{Absolute condition numbers in the coupled radiative energy, with $N=200$, $L_{\rm ang}=6$ and $260$-digit operator and energy arithmetic. Left: the ten selected modes at $(a/M,Q/M,m)=(0.3,0.4,1)$. Blue and gold curves connect the positive- and negative-real-frequency branches, respectively. Red dots mark the selected modes. Small horizontal offsets separate markers within each family. Right: the four modes at $g(\GEapproachT)$, $m=2$. At this offset and for $L_{\rm ang}=6$, both near-EP condition numbers exceed both comparison QNM values (see Sec.~\ref{sec:resolution-tests} for the dependence on offset and truncation). The two pair values are separately tabulated but nearly equal on this scale.}
 \label{fig:ge-conditions}
\end{figure*}

Figure~\ref{fig:ge-conditions} and Table~\ref{tab:ge-coupled-modes} compare the selected modes using the same finite radiative energy. At $(a/M,Q/M,m)=(0.3,0.4,1)$, the ten dominant modes shown belong to the electromagnetic $\ell=1$ family and the gravitational-led and electromagnetic-led $\ell=2,3$ families on the positive- and negative-real-frequency branches. Continuation from the RN limit supports these identifications. Their condition numbers range from $1.55\times10^6$ to $3.54\times10^7$. The full fixed-$m$ operator includes additional modes beyond this selection.

At $g(\GEapproachT)$, both near-EP modes have $\kappa_E\simeq1.97405\times10^{16}$. The two subdominant QNMs have condition numbers $3.03097\times10^{12}$ and $4.10463\times10^{14}$, giving
\begin{equation}
 \frac{\min_{\rm pair}\kappa_E}
      {\max_{\rm subdominant}\kappa_E}\simeq48.09.
\end{equation}
At this offset, and for the $L_{\rm ang}=6$ matrix, each member of the pair is more sensitive than either displayed subdominant mode. This comparison holds the background and discretization fixed and does not imply a universal ordering by damping rate. Section~\ref{sec:resolution-tests} shows that the ordering reverses at larger offsets and that angular refinement displaces the coalescence by more than $t=\GEapproachT$. Comparison with the $m=1$ baseline changes both the background and the angular sector, so it cannot isolate the effect of EP proximity.

\begin{table*}[t]
\centering\small
\begin{tabular}{l@{\quad}l@{\quad}c@{\quad}c}
\hline
Background & Selected mode & $M\omega$ & $\kappa_E$ \\
\hline
Baseline & $\ell=1,\ \mathrm{EM}_{+}$ & $0.2884427876474998-0.0918564166278709i$ & $2.4272\times10^{7}$ \\
Baseline & $\ell=1,\ \mathrm{EM}_{-}$ & $-0.2431846141609501-0.0947193020526147i$ & $1.8328\times10^{7}$ \\
Baseline & $\ell=2,\ G_{+}$ & $0.4048337945835153-0.0882198701113382i$ & $3.5383\times10^{7}$ \\
Baseline & $\ell=2,\ G_{-}$ & $-0.3619980614324120-0.0891536352532961i$ & $2.8496\times10^{7}$ \\
Baseline & $\ell=2,\ \mathrm{EM}_{+}$ & $0.5086472866868466-0.0951389238667152i$ & $5.3671\times10^{6}$ \\
Baseline & $\ell=2,\ \mathrm{EM}_{-}$ & $-0.4603983454513027-0.0960634643217695i$ & $4.1588\times10^{6}$ \\
Baseline & $\ell=3,\ G_{+}$ & $0.6357193722098390-0.0919418050312490i$ & $7.2639\times10^{6}$ \\
Baseline & $\ell=3,\ G_{-}$ & $-0.5902743322275076-0.0926366450677353i$ & $6.1355\times10^{6}$ \\
Baseline & $\ell=3,\ \mathrm{EM}_{+}$ & $0.7173637694369437-0.0958752753450901i$ & $1.9028\times10^{6}$ \\
Baseline & $\ell=3,\ \mathrm{EM}_{-}$ & $-0.6683459517512474-0.0964037144676391i$ & $1.5546\times10^{6}$ \\
\hline
Near EP & near-EP 1 & $0.5643982314875481-0.0283045497842056i$ & $1.9740\times10^{16}$ \\
Near EP & near-EP 2 & $0.5643982330489119-0.0283045476827834i$ & $1.9740\times10^{16}$ \\
Near EP & QNM 1 & $0.5585410513080401-0.0518972401969080i$ & $3.0310\times10^{12}$ \\
Near EP & QNM 2 & $0.5573508543821729-0.0719692905324936i$ & $4.1046\times10^{14}$ \\
\hline
\end{tabular}
\caption{All fourteen selected coupled GE QNMs and absolute condition numbers in the specified energy norm. The baseline has $(a,Q,m)=(0.3,0.4,1)$, while the near-EP case uses $g(\GEapproachT)$ and $m=2$, with the parameter path and unrounded critical candidate specified in the text. All rows use $N=200$, angular cutoff $L_{\rm ang}=6$ and $260$ working digits of precision, including the energy factors. The condition formula includes the mass matrix of the discretized coupled equations and the associated source conversion. Frequency digits distinguish the resolved finite-grid eigenvalues. Condition numbers are displayed with five significant figures. The modes labelled near-EP 1 and near-EP 2 form the nearly coalescing pair, while QNM 1 and QNM 2 label only the two displayed subdominant modes, without assigning global overtone indices.}
\label{tab:ge-coupled-modes}
\end{table*}

We isolate the approach to coalescence by continuing the same $m=2$ pair along Eq.~\eqref{eq:ge-off-ep} at $t=10^{-3},10^{-6},10^{-9},10^{-12}$. The energy definition and relative field normalization remain fixed, with coefficients varying smoothly with the background. Over this path, the gap decreases from $8.27879\times10^{-5}$ to $2.61798\times10^{-9}$. The two condition numbers grow from approximately $6.3\times10^{11}$ and $6.2\times10^{11}$, respectively, to $1.97405\times10^{16}$. On the final interval, the logarithmic slopes are $0.500000$ for the gap and $-0.499998,-0.500002$ for the condition numbers. The four samples agree with the generic square-root splitting and inverse-square-root conditioning expected near an EP at fixed discretization. Accordingly, the product $\kappa_E d\simeq5.17\times10^{7}$ remains constant along the path.

\begin{figure*}[t]
 \centering
 \includegraphics[width=0.99\textwidth]{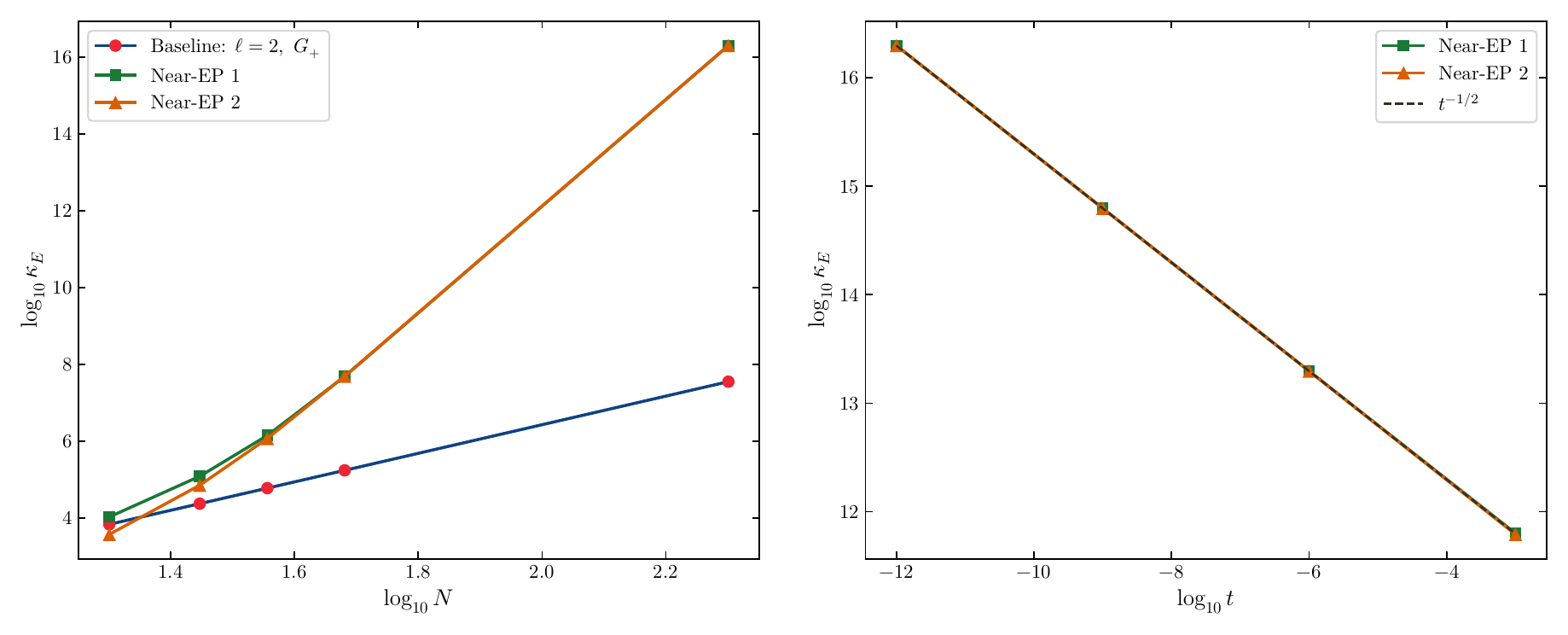}
\caption{Resolution dependence and approach to the coupled finite-grid EP, in the same radiative energy. Left: Positive-branch $\ell=2$ gravitational-led fundamental QNM and the near-EP pair at fixed $g(\GEapproachT)$, for $N=20,28,36,48,200$ and $L_{\rm ang}=6$. The background is tuned to the $N=200$ EP, whose position changes with resolution. Right: the condition numbers of the same pair along the parameter path in Eq.~\eqref{eq:ge-off-ep}. The dashed $t^{-1/2}$ reference is normalized to the closest sample. Both near-EP curves closely follow this reference and overlap at the plotted scale. No offsets are applied to numerical values. All values use $260$ working digits, including the energy factors. The four samples test consistency with generic EP scaling.}
 \label{fig:ge-stability}
\end{figure*}

The resolution survey in Fig.~\ref{fig:ge-stability} uses a fixed, unrounded background. For the positive gravitational-led $\ell=2$ baseline mode, the condition number grows from $6.85\times10^3$ at $N=20$ to $3.54\times10^7$ at $200$, although the frequency changes by only $7.83\times10^{-10}$ between $N=36$ and $48$. The near-EP curves also reflect the displacement of the finite-matrix EP with $N$. Accurate QNM frequencies are already obtained at resolutions for which convergence of the left vectors and resolvent remains unresolved.

\section{Spectral interpretation and physical scope}
\label{sec:discussion}

An individual QNM and a close pair characterize different aspects of spectral instability. The condition number measures the frequency shift of a simple mode under small changes in the equations. The pseudospectrum also describes the collective response of nearby modes, whose merged contours need not be ranked by the largest individual condition number. The matched GE path tests this behavior with the same modes, discretization and normalization rule.

These comparisons depend on how energy is defined. A change of field variables leaves the frequencies, response norm and condition numbers unchanged when the energy matrix and source are transformed consistently. A different choice of energy can change the magnitudes of the response and condition numbers. Near a generic EP, the square-root exponents reflect mode coalescence, while their prefactors depend on the norm.

\subsection{Resolution and truncation tests}
\label{sec:resolution-tests}

As Fig.~\ref{fig:ge-stability} illustrates, convergence of the frequencies alone does not establish convergence of the condition numbers. The latter also depend on the left vectors, energy weights and source conversion. For discretized dissipative BH problems, the pseudospectrum and the condition numbers are not expected to converge with resolution~\cite{Jaramillo2021,Destounis:2021lum}, and we do not interpret their magnitudes as continuum quantities. Only the QNM frequencies, and hence the location of an EP, converge. Higher arithmetic precision reduces roundoff within a chosen discretization but cannot remove radial or angular truncation error. Either truncation can shift the critical background near an EP. The varying-$N$ curves combine resolution effects with a changing distance from the EP selected at $N=200$, $L_{\rm ang}=6$, and cannot be interpreted as a continuum condition-number limit.

We, thus, ask which features of the near-EP comparison persist when the offset, the radial resolution and the angular truncation are varied. The tests use an independent implementation of both discretized problems in binary64 arithmetic, described and validated against the $N=200$ results in Appendix~\ref{app:binary64-checks}. Near-EP roots computed in binary64 carry roundoff of order $10^{-16}\kappa$. The tests therefore rely on quantities that remain well conditioned: the product $\kappa d$, the condition numbers of the simple comparison modes, and first-order changes of the pair discriminant.

Two features persist. First, the EP scaling has an offset-independent prefactor. Near a second-order EP, Eqs.~\eqref{eq:Aminus2-Laurent} and \eqref{eq:pair-projectors} give $\kappa_{1,2}\,d\simeq\|A_{-2}\|_E$ to leading order, so $\kappa d$ characterizes the pair without reference to $t$. For the scalar pair at $N=200$, $\kappa d\simeq660.8$ varies by less than $0.05\%$ between $t=10^{-1}$ and $10^{-6}$. For the GE pair at $N=100$, $\kappa d=8.949\times10^{5}$ to within $10^{-4}$ for $L_{\rm ang}=6,7,8,10$, with and without the factor $f_c$ of Eq.~\eqref{eq:hypkn-clearing-factor}, and at $t=10^{-4}$ and $10^{-3}$. Second, close enough to the EP the pair is more sensitive than any fixed simple mode, since $\kappa_{\rm pair}\propto t^{-1/2}$ while the comparison modes vary smoothly with the background.

\begin{figure*}[t]
 \centering
 \includegraphics[width=0.99\textwidth]{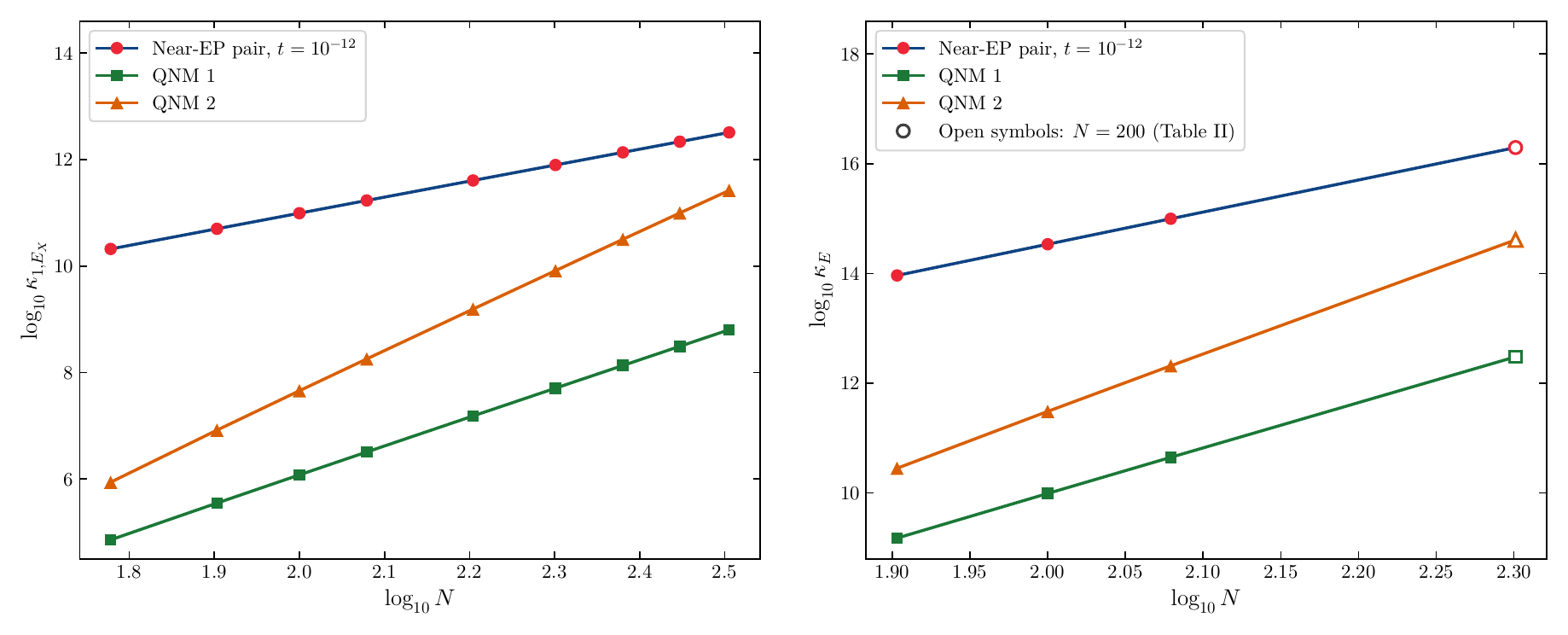}
\caption{Resolution dependence of the near-EP comparison at fixed angular truncation. Left: scalar $\ell=m=1$ branch, $N=60$--$320$. Right: GE, $m=2$, $L_{\rm ang}=6$, $N=80$--$200$. Navy curves show the pair condition number $\kappa d/d(10^{-12})$ at the matched gap of Eqs.~\eqref{eq:scalar-enhanced-background} and \eqref{eq:ge-off-ep}; gold and purple curves show QNM~1 and QNM~2 at $g(10^{-12})$. Filled red dots are computed with the binary64 implementation of Appendix~\ref{app:binary64-checks}. Open circles are the $260$-digit values of Table~\ref{tab:ge-coupled-modes}. The scalar $N=200$ points reproduce Table~\ref{tab:branchconditions}. Values are listed in Table~\ref{tab:resolution-tests}.}
 \label{fig:resolution-ordering}
\end{figure*}

\begin{figure*}[t]
 \centering
 \includegraphics[width=0.99\textwidth]{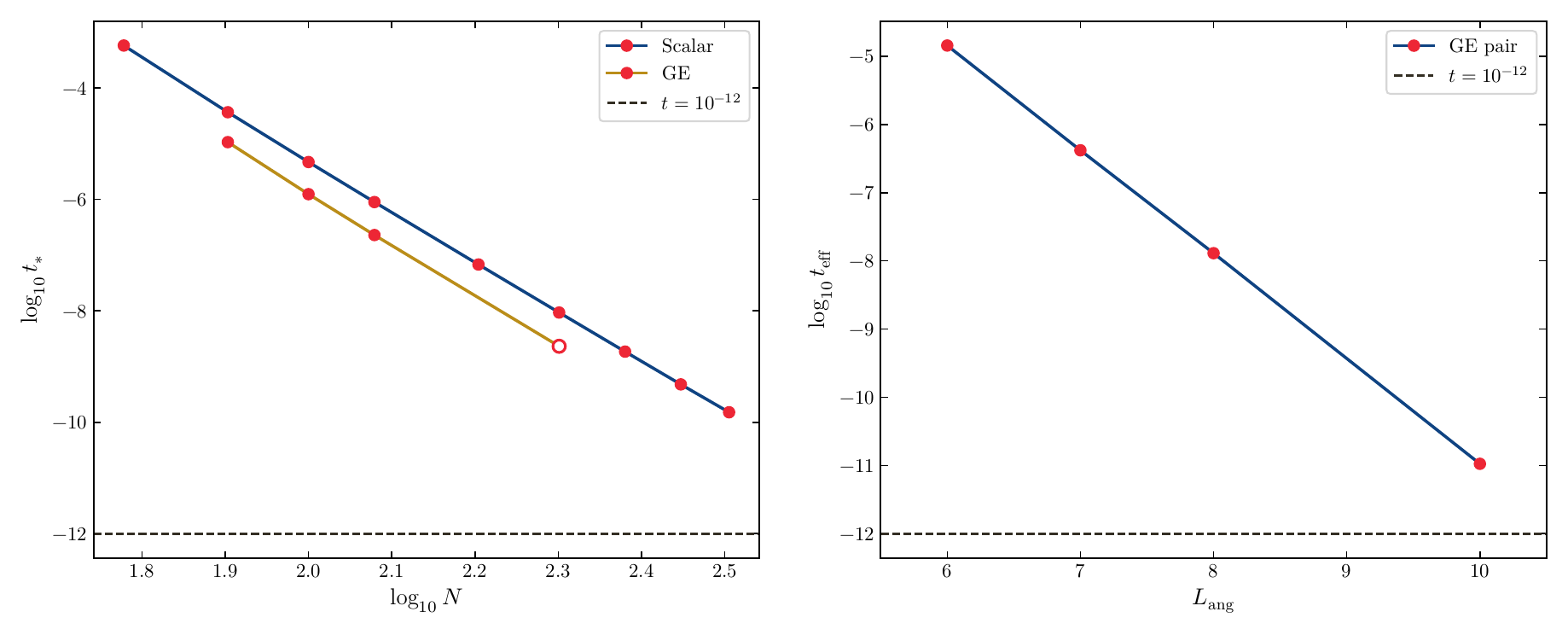}
\caption{Left: crossover offset $t_\ast$ of Eq.~\eqref{eq:crossover-offset}, below which both pair members exceed QNM~2, for the scalar branch and for GE with $L_{\rm ang}=6$. The open circle uses the $260$-digit values of Table~\ref{tab:ge-coupled-modes}. The dashed line marks the offset $t=10^{-12}$ used in Figs.~\ref{fig:scalarbaseline}--\ref{fig:ge-conditions}. Right: equivalent path offset $t_{\rm eff}=|D_{L_{\rm ang}}-D_{12}|/|\partial_tD|$ induced by angular truncation of the GE problem at fixed $(a/M,Q/M)$, where $D=(\omega_1-\omega_2)^2$ is the pair discriminant. The selected offset is resolved only for $L_{\rm ang}\ge12$.}
 \label{fig:crossover-truncation}
\end{figure*}

The ordering at a fixed offset, however, depends on resolution. Table~\ref{tab:resolution-tests} and Fig.~\ref{fig:resolution-ordering} compare the pair, evaluated at the matched gap $d(10^{-12})$, with the comparison modes at the same background. All condition numbers grow with $N$, but the comparison modes grow faster. For the scalar branch, fits over $N\ge100$ give $\kappa d\propto N^{3.0}$, $\kappa({\rm QNM\ 1})\propto N^{5.4}$ and $\kappa({\rm QNM\ 2})\propto N^{7.5}$. For GE over $N=80$--$200$, the corresponding exponents are $5.9$, $8.3$ and $10.4$. The pair-to-QNM~2 ratio at $t=10^{-12}$ falls from $2.4\times10^{4}$ at $N=60$ to $96.5$ at $N=200$ and $12.3$ at $N=320$ for the scalar branch. Extrapolated, it reaches unity near $N\approx560$. For GE, the ratio falls from $3.3\times10^{3}$ at $N=80$ to $48.1$ at $N=200$. The crossover offset below which both pair members exceed QNM~2,
\begin{equation}
 t_\ast=t\left[\frac{\kappa_{\rm pair}(t)}{\kappa({\rm QNM\ 2})}\right]^2,
 \label{eq:crossover-offset}
\end{equation}
decreases approximately as $N^{-9}$ in both problems (Fig.~\ref{fig:crossover-truncation}, left). The qualitative statement that survives is therefore conditional. For every resolution tested, there is an offset below which the pair dominates the displayed modes, but that offset depends on resolution. The factors $96.6$ and $48.09$ quoted in Secs.~\ref{sec:conditioning} and \ref{sec:measured-conditioning} are specific to $N=200$.

\begin{table*}[t]
\centering\small
\begin{tabular}{l@{\quad}r@{\quad}c@{\quad}c@{\quad}c@{\quad}c@{\quad}c}
\hline
Field & $N$ & $\kappa\,d$ & $\kappa$(QNM 1) & $\kappa$(QNM 2) & $\min_{\rm pair}\kappa/\kappa$(QNM 2) & $t_\ast$ \\
\hline
Scalar & 60 & $1.750\times10^{1}$ & $7.188\times10^{4}$ & $8.691\times10^{5}$ & $2.41\times10^{4}$ & $5.8\times10^{-4}$ \\
Scalar & 80 & $4.187\times10^{1}$ & $3.524\times10^{5}$ & $8.253\times10^{6}$ & $6.06\times10^{3}$ & $3.7\times10^{-5}$ \\
Scalar & 100 & $8.217\times10^{1}$ & $1.193\times10^{6}$ & $4.535\times10^{7}$ & $2.17\times10^{3}$ & $4.7\times10^{-6}$ \\
Scalar & 120 & $1.424\times10^{2}$ & $3.214\times10^{6}$ & $1.797\times10^{8}$ & $947$ & $9.0\times10^{-7}$ \\
Scalar & 160 & $3.382\times10^{2}$ & $1.524\times10^{7}$ & $1.551\times10^{9}$ & $261$ & $6.8\times10^{-8}$ \\
Scalar & 200 & $6.608\times10^{2}$ & $5.075\times10^{7}$ & $8.181\times10^{9}$ & $96.5$ & $9.3\times10^{-9}$ \\
Scalar & 240 & $1.142\times10^{3}$ & $1.354\times10^{8}$ & $3.168\times10^{10}$ & $43.1$ & $1.9\times10^{-9}$ \\
Scalar & 280 & $1.811\times10^{3}$ & $3.099\times10^{8}$ & $9.900\times10^{10}$ & $21.9$ & $4.8\times10^{-10}$ \\
Scalar & 320 & $2.702\times10^{3}$ & $6.349\times10^{8}$ & $2.627\times10^{11}$ & $12.3$ & $1.5\times10^{-10}$ \\
\hline
GE & 80 & $2.414\times10^{5}$ & $1.507\times10^{9}$ & $2.821\times10^{10}$ & $3.27\times10^{3}$ & $1.1\times10^{-5}$ \\
GE & 100 & $8.948\times10^{5}$ & $9.795\times10^{9}$ & $3.067\times10^{11}$ & $1.11\times10^{3}$ & $1.2\times10^{-6}$ \\
GE & 120 & $2.605\times10^{6}$ & $4.464\times10^{10}$ & $2.078\times10^{12}$ & $479$ & $2.3\times10^{-7}$ \\
GE & 200$^{\dagger}$ & $5.168\times10^{7}$ & $3.031\times10^{12}$ & $4.105\times10^{14}$ & $48.1$ & $2.3\times10^{-9}$ \\
\hline
\end{tabular}
\caption{Resolution dependence of the near-EP comparison at fixed angular truncation ($\ell\le79$ for the scalar spheroidal branch and $L_{\rm ang}=6$ for GE). The product $\kappa\,d$ uses the geometric mean of the two pair condition numbers at $t=10^{-4}$ and does not depend on $t$ to the digits shown. The comparison modes are evaluated at $g(10^{-12})$. The ratio uses the pair condition number at the matched gap of the $N=200$ calculation, $\kappa_{\rm pair}=\kappa d/d(10^{-12})$, and $t_\ast$ is defined in Eq.~\eqref{eq:crossover-offset}. Scalar condition numbers are $\kappa_{1,E_X}$ of Eq.~\eqref{eq:branchkappa} and GE condition numbers are $\kappa_E$ of Eq.~\eqref{eq:ge-gen-kappa}. All entries except the row marked $^\dagger$ come from the binary64 implementation of Appendix~\ref{app:binary64-checks}. The row marked $^\dagger$ repeats the $260$-digit values of Table~\ref{tab:ge-coupled-modes}. The scalar $N=200$ row reproduces Table~\ref{tab:branchconditions}.}
\label{tab:resolution-tests}
\end{table*}

The GE comparison is also affected by angular truncation, because its background is tuned to the EP of the $L_{\rm ang}=6$ matrix. Differencing nearly coalescing roots between two truncations would lose the effect in roundoff. We instead eliminate the added angular functions exactly through a Schur complement and compute the first-order change of the pair discriminant $D=(\omega_1-\omega_2)^2$, which is analytic in the perturbation even though the individual roots are not. Along Eq.~\eqref{eq:ge-off-ep}, $D\simeq t\,\partial_tD$ with $|\partial_tD|=d^2(10^{-12})/10^{-12}\simeq6.85\times10^{-6}$. A change $\delta D$ at fixed $(a/M,Q/M)$ therefore corresponds to an equivalent offset $t_{\rm eff}=|\delta D|/|\partial_tD|$. Relative to $L_{\rm ang}=12$, Table~\ref{tab:angular-tests} and Fig.~\ref{fig:crossover-truncation} (right) give $t_{\rm eff}=1.4\times10^{-5}$ at $L_{\rm ang}=6$, decreasing geometrically to $1.1\times10^{-11}$ at $L_{\rm ang}=10$. The result is the same at $N=80$ and $100$ and at $t=10^{-4}$ and $10^{-3}$, and it agrees with direct $L_{\rm ang}=7,8,10$ calculations within their roundoff.

In the angularly converged problem, the background $g(\GEapproachT)$ thus lies at an effective offset $t_{\rm eff}\simeq1.4\times10^{-5}$, where the gap is about $1.0\times10^{-5}$ rather than $\GEpairGap$. Equivalently, the coupled EP moves by a few times $10^{-9}$ in $(a/M,Q/M)$, compared with the selected offset $t|g_{\rm ref}-g_{\rm EP}^{(N)}|\simeq5.2\times10^{-18}$. The comparison QNMs move by less than $10^{-9}$ and their condition numbers by less than $0.1\%$. Using the $N=200$ value of $\kappa d$, the pair condition number at $t_{\rm eff}$ is about $5.2\times10^{12}$, which lies between those of QNM~1 and QNM~2. With $L_{\rm ang}=6$, radial truncation at $N=80$--$120$ likewise moves the background, relative to the $N=200$ tuning, by $t_{\rm eff}\sim10^{-6}$--$10^{-5}$. A binary64 calculation cannot resolve the corresponding $N=200$ to $240$ displacement. Placing a GE background at $t=10^{-12}$ from the continuum EP would require retuning with $L_{\rm ang}\ge12$ and a matching radial check. The finite-matrix results of Sec.~\ref{sec:maps}, including the contour merger and the square-root laws, are unaffected, since they refer to the $L_{\rm ang}=6$ matrix and its own EP.

For the scalar branch, the EP location is stable at the binary64 floor. For $60\le N\le200$ and spheroidal cutoffs from $\ell\le7$ to $\ell\le79$, $g_{\rm EP}^{(N)}$ agrees with Eq.~\eqref{eq:scalar-ep-reference} to about $10^{-11}$. This test does not reach the scale $1.2\times10^{-17}$ of the scalar offset, but the spheroidal eigenvalue converges geometrically in the cutoff and the radial frequencies are checked independently in Appendix~\ref{app:scalar-validation}.

\begin{table}[t]
\centering\small
\begin{tabular}{c@{\quad}c@{\quad}c@{\quad}c@{\quad}c}
\hline
$L_{\rm ang}$ & $t_{\rm eff}$ & $|\delta\omega_{\rm QNM\,1}|$ & $|\delta\omega_{\rm QNM\,2}|$ & $\kappa\,d$ \\
\hline
6 & $1.4\times10^{-5}$ & $9.0\times10^{-10}$ & $6.1\times10^{-10}$ & $8.948\times10^{5}$ \\
7 & $4.2\times10^{-7}$ & $2.8\times10^{-11}$ & $2.2\times10^{-11}$ & $8.949\times10^{5}$ \\
8 & $1.3\times10^{-8}$ & $9.4\times10^{-13}$ & $6.8\times10^{-13}$ & $8.949\times10^{5}$ \\
10 & $1.1\times10^{-11}$ & $1.1\times10^{-15}$ & $5.0\times10^{-14}$ & $8.949\times10^{5}$ \\
\hline
\end{tabular}
\caption{Angular truncation of the GE problem near $g(\GEapproachT)$, Eq.~\eqref{eq:ge-off-ep}. The equivalent offset $t_{\rm eff}$ and the comparison-mode shifts $|\delta\omega|$ are measured relative to $L_{\rm ang}=12$, at $N=80$--$100$, by exact Schur elimination of the added angular functions and first-order perturbation. They are identical at $g(10^{-4})$ and $g(10^{-3})$ to the digits shown. The product $\kappa\,d$ uses $N=100$ and $t=10^{-4}$.}
\label{tab:angular-tests}
\end{table}

\subsection{The outgoing-wave problem}

The outgoing-wave spectral problem also requires endpoint conditions beyond a positive energy integral. An unwanted solution may decay so rapidly at a compactified endpoint that ordinary smoothness tests admit it. Specifying the allowed endpoint behavior is essential in this setting. Hyperboloidal resonance expansions and rigorous Kerr results clarify the issue~\cite{BessonJaramillo2025,GajicWarnick2024}, but the corresponding coupled KN construction is still needed. Appendix~\ref{sec:domain} describes the endpoints for the fields used here. The figures represent the discretized response, with separate frequency, energy, arithmetic and interpolation checks reported in Appendices~\ref{app:scalar-validation} and \ref{app:numerics}.

\subsection{Meaning of the radiative energy}

The scalar norm follows from the field's stress-energy tensor, whereas the GE norm assigns a positive energy to the chosen curvature and connection fields. Their relative weights and normalization determine which field changes count as small. Relating these changes to physical Einstein--Maxwell perturbations requires satisfying the constraints and reconstructing the metric and electromagnetic variables. The GE norm used here has not been identified with the canonical energy of those perturbations or with the energy needed to produce an arbitrary physical source.

Relevant approaches include positive energies for axisymmetric KN perturbations~\cite{MoncriefGudapati2023}, coupled physical-space estimates~\cite{Giorgi2020,Giorgi2023}, canonical-energy methods~\cite{HollandsWald2013} and radiative-amplitude reconstruction~\cite{Motohashi2026}. Each uses different states or assumptions. To obtain a physical gravitational or electromagnetic signal from the response studied here, one must supply the appropriate source and extraction maps. Until that calibration is available, the absolute areas of scalar and GE contours cannot be interpreted as a common physical energy cost.

The coupled finite-generator condition numbers measure the largest sensitivity to arbitrary matrix perturbations in the chosen norm. A change in a physical parameter samples a restricted class of perturbations and may produce a smaller frequency shift. For the stored problem $K_Nv_j=\omega_jM_Nv_j$, a simple QNM satisfies
\begin{equation}
 \frac{\partial\omega_j}{\partial p}=
 \frac{w_j^\dagger(\partial_pK_N-\omega_j\partial_pM_N)v_j}
 {w_j^\dagger M_Nv_j},\qquad p\in\{a,Q\}.
 \label{eq:structured-parameter-sensitivity}
\end{equation}
Calculating these parameter derivatives and the response to sources satisfying the Einstein--Maxwell constraints would establish how much of the measured sensitivity is accessible to physical perturbations. Both calculations lie beyond the scope of the present analysis.

\section{Conclusions and outlook}
\label{sec:conclusions-outlook}

Our calculations examine energy-norm spectral instability near EPs in neutral massless scalar and coupled GE KN perturbations. Along matched paths toward the selected EPs, the frequency gap and condition numbers follow the expected square-root and inverse-square-root laws. Their product, which approximates $\|A_{-2}\|_E$, does not depend on the offset or, for GE, on the angular truncation and projection prescription. At the selected backgrounds and $N=200$, both members of each close pair have larger condition numbers than either of the two displayed comparison modes, by factors of about $96.6$ in the scalar case and $48.09$ in the GE case. These factors depend on resolution. The comparison modes become ill conditioned faster with $N$ than the pair, so the pair dominates only below a crossover offset that decreases approximately as $N^{-9}$. For the GE pair, angular refinement beyond $L_{\rm ang}=6$ also displaces the coalescence by more than the selected offset, and at fixed parameters in the angularly converged problem the pair lies between the two comparison modes. The contours and condition numbers are therefore consistent with a local amplification of spectral sensitivity near EPs in these numerical problems. Its strength relative to other modes depends on the distance to the EP and on the discretization, and we do not infer a universal ordering of QNMs or overtones.

The collective response may be relevant to BH spectroscopy. At a second-order EP of the generator, the coalesced pair contributes
\begin{equation}
 e^{-iL_{\rm EP}T}\Pi
 =e^{-i\omega_{\rm EP}T}
  \bigl(\Pi-iT A_{-2}\bigr),
 \label{eq:outlook-EP-propagator}
\end{equation}
where $\Pi$ selects the full two-mode subspace and need not be orthogonal in the chosen energy. The matrix $A_{-2}$ is the double-pole coefficient defined in Sec.~\ref{sec:framework}. Let $\mathcal I$ and $\mathcal O$ be suitable time-independent maps from physical initial data to the evolution fields and from those fields to the waveform, respectively. If these maps are available, initial data $\xi$ would produce the contribution
\begin{equation}
 \begin{aligned}
 h_{\rm EP}(T)&=e^{-i\omega_{\rm EP}T}(h_0+T h_1),\\
 h_0&=\mathcal O\Pi\mathcal I\xi,\\
 h_1&=-i\mathcal O A_{-2}\mathcal I\xi.
 \end{aligned}
 \label{eq:outlook-EP-waveform}
\end{equation}
This time dependence is characteristic of exact mode coalescence and agrees with EP-based ringdown and sourced-response descriptions~\cite{YangBertiFranchini2025,PanossoMacedo:2025xnf,Morikawa:2026fzn}. The coefficient of the linear term can vanish for particular excitation or observation channels. Large individual modal contributions may also interfere destructively, leaving a finite collective waveform~\cite{KNOutlookOshita2025}. A large energy-norm condition number by itself predicts neither the observed strain nor its detectability. For $\operatorname{Im}\omega_{\rm EP}<0$, the polynomial prefactor is compatible with eventual exponential decay. A complementary gravitational-wave scattering calculation would require physical incident and outgoing amplitudes at real frequencies, which are not determined by the complex-frequency contours alone~\cite{Torres:2023nqg,Rosato:2024arw}.

The selected GE pair illustrates the difficulty of inferring two nearly coincident modes. Its central frequency is $\omega_c=(\omega_1+\omega_2)/2$, and its dimensionless mean damping rate is $\gamma=-\operatorname{Im}(M\omega_c)>0$. The corresponding amplitude damping time is $\tau=M/\gamma$. For the dimensionless complex-frequency separation $d=|M(\omega_1-\omega_2)|$, the reported values give
\begin{equation}
 \frac{d}{\gamma}\simeq9.25\times10^{-8}.
 \label{eq:outlook-gap-damping}
\end{equation}
Over one damping time, the two exponentials acquire very little differential evolution. A collective pair description that remains finite at coalescence may be more suitable than independent fits of two modal amplitudes. This comparison of timescales is heuristic and does not define a resolution threshold. Simulated signals with detector noise could test recovery of the pair center and splitting using ordinary ringdown models and regular two-mode models~\cite{BertiCardosoWill2006,YangBertiFranchini2025,KubotaMotohashi2026}. Such tests should include additional QNMs, the prompt response and tails where relevant~\cite{Leaver1986}.

Another question is how a gravitational-wave response changes when the background evolves near an EP. Because an EP is a degeneracy in the parameter space of the evolution operator, neither spatial propagation of a wave packet nor a sweep of the real driving frequency on a stationary KN background takes the background through it. A passage would require a physically admissible history $g(T)=(a/M,Q/M)$ approaching the critical parameters. The stationary path parameter in this work is not physical time. Absorption-driven changes in a Kerr BH's mass and spin can already alter mode excitation without an EP passage~\cite{KNOutlookMay2024}. A KN calculation would require consistent flux balance and charge conservation, together with the additional terms introduced by time-dependent coordinates and field normalizations. Even at conserved electric charge $Q$, the ratio $Q/M$ can change through the evolution of $M$.

An EP in a generic complex non-Hermitian family requires two real tuning conditions~\cite{KNOutlookMailybaev2005}. An exact crossing along a one-parameter history would require corresponding tuning. Even a nearby passage calls for an assessment of the coupling induced by background evolution before replacing the dynamics with QNMs of successive stationary BHs~\cite{UzdinMailybaevMoiseyev2011}. Direction-dependent mode conversion occurs in other non-Hermitian dynamical systems~\cite{KNOutlookMilburn2015,KNOutlookDoppler2016}, but has yet to be established for the KN gravitational-wave response. Exchange of stationary eigenvalue branches, dynamical state transfer and hysteresis are distinct phenomena. The stationary pseudospectra calculated here do not demonstrate the latter two effects. Their astrophysical relevance would also depend on whether admissible histories approach the charged near-extremal backgrounds considered in this work.

Further work should establish the outgoing-wave continuum problem and retune the GE EP with $L_{\rm ang}\ge12$ and increasing $N$. Near-EP and comparison modes should then be compared at offsets chosen relative to the resolution-dependent crossover, with resolution-robust measures such as $\kappa d$ preferred to absolute condition numbers. Physical sources and gravitational/electromagnetic waveform extraction can then be used to determine whether the near-EP spectral sensitivity found here leaves a measurable collective ringdown signature.

\begin{acknowledgments}
K.D. acknowledges financial support provided by FCT--Funda\c{c}\~ao para a Ci\^encia e a Tecnologia, I.P., under the Scientific Employment Stimulus -- Individual Call -- Grant No. 2023.07417.CEECIND/CP2830/CT0008. K.D. would also like to thank the Funda\c{c}\~ao para a Ci\^encia e Tecnologia (FCT), Portugal, for the financial support to the Center for Astrophysics and Gravitation (CENTRA/IST/ULisboa) through grant No. UID/PRR/00099/2025 and grant No. UID/00099/2025.
This project has received funding from the European Union's Horizon MSCA-2022 research and innovation programme ``Einstein Waves'' under grant agreement No. 101131233.
The scientific conception and research direction of this work originated with the author. ChatGPT 6.1 Sol was used under the author's direction to assist with manuscript drafting and revision, and selected aspects of figure presentation. Claude (Anthropic) was used under the author's direction to implement independent binary64 checks of Sec.~\ref{sec:resolution-tests} and Appendix~\ref{app:binary64-checks}. The author reviewed and revised the AI-assisted material and takes responsibility for the final manuscript.
\end{acknowledgments}

\appendix
\section{Hyperboloidal geometry and scalar-field equations}
\label{app:hypkn}
\label{app:geometry}

We construct spacelike slices that cross the future event horizon and reach future null infinity, then derive the scalar wave equation and its positive energy in the resulting coordinates. Appendix~\ref{app:scalar-validation} describes the scalar discretization, and Appendix~\ref{app:ge-energy} develops the coupled GE equations and energy. For the underlying hyperboloidal construction, see Refs.~\cite{HypKN_Zenginoglu2011,HypKN_AssaadMacedo2025}. Its application to BH pseudospectra is discussed in Ref.~\cite{Jaramillo2021}.

\subsection{Conventions and background geometry}
\label{app:hypkn-background}

Throughout this appendix, $G=c=M=1$, the metric signature is $(-,+,+,+)$, and the frequency convention is $e^{-izT}$, with $z=M\omega_{\rm phys}$. Physical time is related by $T_{\rm phys}=MT$, and damped modes have $\operatorname{Im}z<0$. We distinguish the Boyer--Lindquist azimuth $\phi$ from the horizon-regular azimuth $\varphi$. For the metric and background functions in Eqs.~\eqref{eq:knx-metric} and \eqref{eq:knx-functions}, with $M=1$, define
\begin{equation}
 A_r=r^2+a^2.
 \label{eq:hypkn-background-functions}
\end{equation}
The horizon radii satisfy
\begin{align}
 R&=r_+=1+\sqrt{1-a^2-Q^2},\nonumber\\
 r_-&=1-\sqrt{1-a^2-Q^2},\nonumber\\
 c_0&=a^2+Q^2=Rr_-,\qquad d_H=R-r_->0.
 \label{eq:hypkn-horizons}
\end{align}
We restrict attention to subextremal backgrounds, $a^2+Q^2<1$, and the exterior $r\geq R$. The horizon angular velocity and surface gravity are
\begin{equation}
 \Omega_H=\frac{a}{R^2+a^2},\qquad
 \kappa_H=\frac{d_H}{2(R^2+a^2)}.
 \label{eq:hypkn-horizon-parameters}
\end{equation}
The subscript distinguishes the surface gravity $\kappa_H$ from the mode condition number $\kappa_E$.

\subsection{Radial fixing and the height function}
\label{app:hypkn-coordinates}

Define the tortoise coordinate and azimuthal shift through
\begin{equation}
 \frac{dr_*}{dr}=\frac{A_r}{\Delta},\qquad
 \frac{dk}{dr}=\frac{a}{\Delta}.
 \label{eq:hypkn-tortoise}
\end{equation}
Convenient antiderivatives are
\begin{align}
 r_*={}&r+\frac{R^2+a^2}{d_H}\ln(r-R)\nonumber\\
       &-\frac{r_-^2+a^2}{d_H}\ln(r-r_-)+C_*,\nonumber\\
 k={}&\frac{a}{d_H}\ln\!\left(\frac{r-R}{r-r_-}\right)+C_k.
 \label{eq:hypkn-antiderivatives}
\end{align}
The constants absorb the reference lengths in the logarithms. The advanced coordinates $v=t+r_*$ and $\varphi=\phi+k$ are regular at the future horizon.

We compactify the exterior with
\begin{align}
 x&=\frac{R}{r},\qquad 0\leq x\leq1,\nonumber\\
 T&=t+\mathfrak h(r)=v+4\ln x-2r,\nonumber\\
 \mathfrak h(r)&=r_*+4\ln(R/r)-2r,\nonumber\\
 \varphi&=\phi+k(r).
 \label{eq:hypkn-coordinates}
\end{align}
The numerical implementation uses $T_{\rm code}=t+r_*-2r-4\ln r$, which differs only by the constant shift $T=T_{\rm code}+4\ln R$. Frequencies at a fixed background are unchanged. In these coordinates the event horizon is fixed at $x=1$ and null infinity at $x=0$. The inner horizon, at $x_-=R/r_->1$ when $r_->0$, is not fixed. A Cauchy-horizon fixing instead uses
\begin{equation}
 \sigma=\frac{R-r_-}{r-r_-}
       =\frac{(R-r_-)x}{R-r_-x},
 \label{eq:hypkn-Cauchy-alternative}
\end{equation}
which sends the inner horizon to infinity. The two choices can have different resolution requirements near extremality \cite{Destounis:2021lum,HypKN_AssaadMacedo2025}.

Set $b(r)=2+4/r$. Then
\begin{align}
 dv&=dT+b\,dr,\qquad
 \mathfrak h'=\frac{A_r}{\Delta}-b,\nonumber\\
 \partial_t&=\partial_T,\qquad \partial_\phi=\partial_\varphi,\nonumber\\
 \left.\partial_r\right|_{t,\phi}
 &=-\frac{x^2}{R}\partial_x+\left(\frac{A_r}{\Delta}-b\right)\partial_T
   +\frac{a}{\Delta}\partial_\varphi.
 \label{eq:hypkn-chain-rule}
\end{align}
The second-derivative transformation also differentiates the coefficients in the last line.

At the horizon, $T-v\to-2R$. At infinity, $r_*=r+2\ln r+C_*+O(r^{-1})$, and hence
\begin{equation}
 T-(t-r_*)=2C_*+4\ln R+O(r^{-1}).
 \label{eq:hypkn-time-asymptotics}
\end{equation}
The slices cross the future horizon at finite advanced time and approach future null infinity at finite retarded time. In terms of $\tau=T/R$, this choice is equivalently $v=R[\tau-h(x)]$, with $h(x)=4\ln x/R-2/x$~\cite{MisasArcos2026}.

\subsection{Regular geometry and conformal infinity}
\label{app:hypkn-geometry}

A useful horizon-regular form of the metric is
\begin{align}
 ds^2={}&-dv^2+2\,dv\,dr
       -2a\sin^2\theta\,dr\,d\varphi\nonumber\\
       &+\Sigma\,d\theta^2+A_r\sin^2\theta\,d\varphi^2\nonumber\\
       &+\frac{2r-Q^2}{\Sigma}
       (dv-a\sin^2\theta\,d\varphi)^2.
 \label{eq:hypkn-ingoing-metric}
\end{align}
The substitution $dv=dT+b\,dr$ introduces no inverse powers of $\Delta$, making regularity at $\mathcal H^+$ manifest.

Define
\begin{align}
 C(r)&=A_r-b\Delta\nonumber\\
 &=-r^2+8-a^2-2Q^2-\frac{4c_0}{r},\nonumber\\
 H(x,\theta)&=2A_rb-b^2\Delta-a^2\sin^2\theta\nonumber\\
 &=16-4Q^2-a^2\sin^2\theta\nonumber\\
 &\quad+\frac{(32-8a^2-16Q^2)x}{R}
       -\frac{16c_0x^2}{R^2},\nonumber\\
 K(x)&=-\frac{x^2C}{R},\nonumber\\
 J(x)&=-a\left(1+\frac{4x}{R}\right),\nonumber\\
 f(x)&=1-\frac{2x}{R}+\frac{c_0x^2}{R^2}.
 \label{eq:hypkn-geometric-coefficients}
\end{align}
The nonzero independent inverse-metric components and determinant are
\begin{align}
 \Sigma g^{TT}&=-H,&\Sigma g^{Tx}&=K,\nonumber\\
 \Sigma g^{T\varphi}&=J,&\Sigma g^{xx}&=x^2f,\nonumber\\
 \Sigma g^{x\varphi}&=-\frac{ax^2}{R},&\Sigma g^{\theta\theta}&=1,\nonumber\\
 \Sigma g^{\varphi\varphi}&=\csc^2\theta,&
 \sqrt{-g}&=\frac{\Sigma R\sin\theta}{x^2}.
 \label{eq:hypkn-inverse-metric}
\end{align}
Regular angular fields remove the coordinate singularities at the polar axes.

A $T$-slice is spacelike when $g^{TT}<0$. Since $g^{TT}=-H/\Sigma$, it suffices to establish positivity. At fixed $\theta$, $H$ is a concave quadratic in $1/r$, whose minimum on $0\leq1/r\leq1/R$ occurs at an endpoint. At those endpoints,
\begin{align}
 H_\infty&=16-4Q^2-a^2\sin^2\theta>12,\nonumber\\
 H_H&=2(R^2+a^2)(2+4/R)\nonumber\\
 &\quad-a^2\sin^2\theta>12.
 \label{eq:hypkn-spacelike}
\end{align}
These inequalities establish $H>0$ throughout the exterior, so every slice is spacelike.

With the ADM convention
\begin{equation*}
 ds^2=-\alpha^2dT^2+
 \gamma_{ij}(dx^i+\beta^idT)(dx^j+\beta^jdT),
\end{equation*}
the lapse, shift, normal, and spatial volume element are
\begin{align}
 \alpha^2&=\frac{\Sigma}{H},\qquad
 \beta^x=\frac{K}{H},\nonumber\\
 \beta^\theta&=0,\qquad\beta^\varphi=\frac{J}{H},\nonumber\\
 n&=\alpha^{-1}
 (\partial_T-\beta^x\partial_x-\beta^\varphi\partial_\varphi),\nonumber\\
 \sqrt{\gamma}&=\frac{R\sqrt{\Sigma H}\sin\theta}{x^2}.
 \label{eq:hypkn-ADM}
\end{align}
The inverse spatial metric follows from $\gamma^{ij}=g^{ij}+\beta^i\beta^j/\alpha^2$.

For the conformal factor $\Omega=1/r=x/R$, set $\widetilde g=\Omega^2g$. Its limiting line element at $x=0$ is
\begin{align}
 d\widetilde s^2\big|_{x=0}={}&\frac{2}{R}\,dT\,dx
 +\frac{16-4Q^2}{R^2}\,dx^2\nonumber\\
 &+\frac{2a\sin^2\theta}{R}\,dx\,d\varphi\nonumber\\
 &+d\theta^2+\sin^2\theta\,d\varphi^2.
 \label{eq:hypkn-conformal-metric}
\end{align}
Away from the polar coordinate singularities, this metric is non-degenerate. The limits
\begin{align}
 \widetilde g^{ab}\nabla_a\Omega\nabla_b\Omega
 &=\frac{\Omega^2\Delta}{\Sigma}\longrightarrow0,\nonumber\\
 \Omega\alpha&\longrightarrow H_\infty^{-1/2}.
 \label{eq:hypkn-null-scri}
\end{align}
identify $x=0$ as future null infinity, with generator $\partial_T$.

\subsection{Radial propagation and boundary behavior}
\label{app:hypkn-domain}

For radial propagation, setting the angular wave numbers to zero gives the characteristic speeds
\begin{align}
 \lambda_\pm&=\frac{-K\pm\sqrt{K^2+Hx^2f}}{H}\nonumber\\
 &=\frac{x^2}{RH}\left(C\pm\sqrt{\mathcal A}\right).
 \label{eq:hypkn-characteristics}
\end{align}
The identity $C^2+H\Delta=\mathcal A$ makes the two expressions equivalent, with the signs labeling the same pair of speeds. At $x=0$ the speeds are $0$ and $-2R/H_\infty$. At $x=1$ they are $0$ and $2(R^2+a^2)/(RH_H)$. Both boundaries have one outward and one tangential characteristic. There are no freely specifiable incoming data, but the equations impose compatibility conditions at each endpoint.

The physical scalar phases are ingoing at the future horizon and outgoing at infinity:
\begin{align}
 R_{\rm BL}&\sim e^{-i(z-m\Omega_H)r_*}\quad(r\to R),\nonumber\\
 R_{\rm BL}&\sim r^{-1}e^{+izr_*}\quad(r\to\infty).
 \label{eq:hypkn-scalar-phases}
\end{align}
The rescaling below removes these phases from the unknown radial field. Outgoing solutions must also satisfy the appropriate boundary regularity, which is a stronger requirement than finite energy in some frequency ranges. Appendix~\ref{sec:domain} discusses this distinction.

\subsection{Scalar equation and its first-order generator}
\label{app:hypkn-scalar}

For a neutral massless scalar, $\Box_g\Phi=0$. Multiplication by $\Sigma$ gives the Boyer--Lindquist equation
\begin{align}
 0={}&-\left(\frac{A_r^2}{\Delta}-a^2\sin^2\theta\right)\Phi_{tt}\nonumber\\
 &-\frac{2a(2r-Q^2)}{\Delta}\Phi_{t\phi}
       +\partial_r(\Delta\partial_r\Phi)\nonumber\\
 &+\frac{1}{\sin\theta}\partial_\theta
       (\sin\theta\partial_\theta\Phi)\nonumber\\
 &+\left(\csc^2\theta-\frac{a^2}{\Delta}\right)\Phi_{\phi\phi}.
 \label{eq:hypkn-scalar-BL}
\end{align}
Introduce the rescaled radiation field and its azimuthal decomposition,
\begin{align}
 \Phi&=\frac{e^{im\varphi}}{r\sqrt{2\pi}}\,u(T,x,\theta),\nonumber\\
 D_m&=\partial_\theta^2+\cot\theta\,\partial_\theta
          -m^2\csc^2\theta.
 \label{eq:hypkn-scalar-field}
\end{align}
The azimuthal normalization agrees with the main text. Adding the complex-conjugate azimuthal sector gives a real field.

Let $d_0=8-a^2-2Q^2$ and write
\begin{align}
 H&=H_0(x)+a^2\cos^2\theta,
 \nonumber\\
 H_0&=16-4Q^2-a^2\nonumber\\
 &\quad+\frac{(32-8a^2-16Q^2)x}{R}
    -\frac{16c_0x^2}{R^2},
 \nonumber\\
 K&=R-\frac{d_0x^2}{R}+\frac{4c_0x^3}{R^2},
 \nonumber\\
 E_m&=-2ima-\frac{(2d_0+8ima)x}{R}
                    +\frac{12c_0x^2}{R^2},
 \nonumber\\
 B_m&=2x-\frac{(6+2ima)x^2}{R}
                    +\frac{4c_0x^3}{R^2},
 \nonumber\\
 V_m&=-\frac{2(1+ima)x}{R}+\frac{2c_0x^2}{R^2},
 \nonumber\\
 \mathcal S_m&=x^2f\,\partial_x^2+B_m\partial_x\nonumber\\
 &\quad+D_m+V_m.
 \label{eq:hypkn-scalar-coefficients}
\end{align}
Equation~\eqref{eq:hypkn-chain-rule} then reduces the wave equation to
\begin{equation}
 -H\,u_{TT}+2K\,u_{Tx}+E_m u_T+\mathcal S_m u=0.
 \label{eq:hypkn-scalar-hyperboloidal}
\end{equation}
All coefficients have regular radial limits, including the angular contribution $a^2\cos^2\theta$ to $H$.

Define the independent velocity variable $v=i\partial_Tu$ and $U=(u,v)^T$. Then
\begin{align}
 i\partial_TU&=\mathcal L_mU,\nonumber\\
 \mathcal L_m&=\begin{pmatrix}
 0&I\\-H^{-1}\mathcal S_m&iH^{-1}\mathcal T_m
 \end{pmatrix},\nonumber\\
 \mathcal T_m&=2K\partial_x+E_m.
 \label{eq:hypkn-scalar-generator}
\end{align}
Multiplication by $H^{-1}$ acts after the derivatives to its right. The field and velocity are independent components of the evolution state. The relation $v=zu$ applies only to a mode.

In the Schwarzschild limit, $a=Q=0$, we recover $R=2$ and
\begin{align}
 H&=16(1+x),\qquad K=2-4x^2,\nonumber\\
 E_m&=-8x,\nonumber\\
 \mathcal S_m&=x^2(1-x)\partial_x^2\nonumber\\
 &\quad+(2x-3x^2)\partial_x+D_m-x.
 \label{eq:hypkn-Schwarzschild-check}
\end{align}
This limit checks the time convention and the sign of the mixed derivative term.

\subsection{Scalar separation and radial equation}
\label{app:hypkn-scalar-separation}

For $u=e^{-izT}U_{\ell m}(x)S_{\ell m}(\theta;az)$, the angular equation is
\begin{align}
 [D_m+a^2z^2\cos^2\theta+A_{\ell m}(az)]S_{\ell m}&=0,\nonumber\\
 A_{\ell m}(0)&=\ell(\ell+1).
 \label{eq:hypkn-angular-scalar}
\end{align}
We follow the branch that reduces to the regular spherical harmonic at $az=0$. Regularity at the poles requires $S_{\ell m}\sim\sin^{|m|}\theta$. For complex $z$, the angular equation is generally non-Hermitian.

The separated radial equation is
\begin{align}
 \mathcal P_{\ell m}(z)U_{\ell m}&=0,
 \nonumber\\
 \mathcal P_{\ell m}(z)
 &=p_2\partial_x^2+p_1\partial_x+p_0,
 \nonumber\\
 p_2&=x^2f,
 \nonumber\\
 p_1&=B_m-2izK,
 \nonumber\\
 p_0&=V_m-izE_m+z^2H_0-A_{\ell m}(az).
 \label{eq:hypkn-scalar-radial}
\end{align}
Its Boyer--Lindquist counterpart is
\begin{align}
 0={}&\partial_r(\Delta\partial_r R_{\rm BL})\nonumber\\
 &+\biggl[\frac{(A_rz-am)^2}{\Delta}-a^2z^2\nonumber\\
 &\qquad+2amz-A_{\ell m}(az)\biggr]R_{\rm BL},\nonumber\\
 R_{\rm BL}(r)={}&r^{-1}e^{-iz\mathfrak h(r)+imk(r)}U_{\ell m}(R/r).
 \label{eq:hypkn-scalar-radial-BL}
\end{align}
The transformation recovers both physical phases in Eq.~\eqref{eq:hypkn-scalar-phases} while avoiding the exponentially large endpoint amplitudes of the Boyer--Lindquist description.

Since $p_2(0)=p_2(1)=0$, a regular solution satisfies the boundary equations
\begin{align}
 p_1(x_b)U'_{\ell m}(x_b)+p_0(x_b)U_{\ell m}(x_b)&=0,\nonumber\\
 x_b&=0,1.
 \label{eq:hypkn-scalar-endpoints}
\end{align}
In particular,
\begin{align}
 p_1(0)&=-2izR,
 \nonumber\\
 p_0(0)&=z^2(16-4Q^2-a^2)-2amz-A_{\ell m}(az),
 \nonumber\\
 p_1(1)&=\frac{-(2R-2)-2ima+2iz(R^2+a^2)}{R}.
 \label{eq:hypkn-scalar-endpoint-coefficients}
\end{align}
These are endpoint limits of the differential equation, rather than reflecting boundary conditions. A vanishing leading coefficient requires a separate examination of the corresponding endpoint equation.

\subsection{Positive scalar energy}
\label{app:hypkn-scalar-energy}

Starting from the stress-energy tensor, multiplier $X$ and energy flux $E_X$ in Sec.~\ref{sec:scalar}, we now express the energy in compact coordinates. The slice normal $n$, given by Eq.~\eqref{eq:hypkn-ADM}, differs from the locally nonrotating time direction $X$.

The following regular functions are useful for evaluating the energy:
\begin{align}
 \overline\Sigma&=1+\frac{a^2x^2}{R^2}\cos^2\theta,
 \nonumber\\
 W&=\left(1+\frac{a^2x^2}{R^2}\right)^2
             -\frac{a^2x^2}{R^2}f\sin^2\theta,
 \nonumber\\
 \Omega_Z&=\frac{ax^3(2R-Q^2x)}{R^4W},
 \nonumber\\
 \Gamma_X&=\frac{ax}{RW}\biggl[\overline\Sigma\nonumber\\
 &\quad+\left(2+\frac{4x}{R}\right)
       \left(\frac{2x}{R}-\frac{Q^2x^2}{R^2}\right)\biggr].
 \label{eq:hypkn-energy-coefficients}
\end{align}
Completing squares and integrating over $\varphi$ gives
\begin{align}
 E_X[U]&=\frac{1}{2R}\int_0^1\!dx\int_0^\pi\!d\theta\,
                   \sin\theta\,\mathcal E_X,\nonumber\\
 \mathcal E_X&=H|v-m\Omega_Zu|^2\nonumber\\
 &\quad+f|u+x\partial_xu-im\Gamma_Xu|^2\nonumber\\
 &\quad+|\partial_\theta u|^2
       +\frac{m^2\overline\Sigma^2}{W\sin^2\theta}|u|^2.
 \label{eq:hypkn-scalar-compact-energy}
\end{align}
The radial coefficients are finite at both endpoints, and the angular factors act on regular fields. The prefactor follows from the $1/\sqrt{2\pi}$ normalization in Eq.~\eqref{eq:hypkn-scalar-field}. Changing this common normalization for states and residuals leaves their norm ratio unchanged. The four nonnegative terms establish positivity for nonzero states in the $\ell=m=1$ sector. Appendix~\ref{app:scalar-validation} gives the expanded density used in the numerical quadrature.

In an ADM derivation, write $X=\alpha n+S$ with $S^i=\beta^i+\Omega_Z\delta^i_{\varphi}$, and $\Pi=n^a\nabla_a\Phi$. The density takes the form
\begin{align}
 T(X,n)={}&\frac{\alpha}{2}
       \bigl(|\Pi|^2+\gamma^{ij}\partial_i\Phi^*\partial_j\Phi\bigr)\nonumber\\
       &+\operatorname{Re}(\Pi^*S^i\partial_i\Phi).
 \label{eq:hypkn-energy-ADM-derivation}
\end{align}
Substituting the compact variables into this density gives Eq.~\eqref{eq:hypkn-scalar-compact-energy}. To check the square completion, define
\begin{align}
 G_X&=\frac{x}{R}(a-\Omega_ZC),\nonumber\\
 D_X&=\csc^2\theta-2\Omega_ZJ-H\Omega_Z^2.
 \label{eq:hypkn-energy-completion}
\end{align}
Then $G_X=f\Gamma_X$ and $D_X-f\Gamma_X^2=\overline\Sigma^2/(W\sin^2\theta)$. The first identity completes the radial square. The second gives the remaining angular coefficient in Eq.~\eqref{eq:hypkn-scalar-compact-energy}.

For solutions, $\nabla_a(T^{ab}X_b)=T^{ab}\nabla_{(a}X_{b)}$. Between two slices, including the horizon and null-infinity fluxes, the energy balance is
\begin{align}
 E_X(T_2)+\mathcal F_{\mathcal H^+}+\mathcal F_{\mathscr I^+}
 ={}&E_X(T_1)\nonumber\\
 &-\int_{\mathcal V}T^{ab}\nabla_{(a}X_{b)}\,dV.
 \label{eq:hypkn-scalar-balance}
\end{align}
The outgoing fluxes use the positive physical-energy orientation of the current $-T^a{}_bX^b$. Because $\Omega_Z$ varies in space, the bulk term need not vanish. The energy is positive but need not be conserved or decrease monotonically.
\section{Scalar discretization and validation}
\label{app:scalar-validation}

\subsection{Radial grid and coefficients}
\label{app:scalar-equations}

We discretize the separated equations of Appendix~\ref{app:hypkn} in the coefficient form given below. The scalar and GE calculations share the radial nodes
\begin{equation}
 x_j=\frac12\left[1+\cos\!\left(\frac{j\pi}{N-1}\right)\right],
 \quad j=0,\ldots,N-1.
 \label{eq:hypkn-radial-grid}
\end{equation}
The interpolating polynomial has degree $N-1$ on these $N$ nodes. Set $c=a^2+Q^2$, $\beta=8-a^2-2Q^2$ and $d(x)=1-2x/R+cx^2/R^2$. The operators in Eq.~\eqref{eq:scalar-companion} are
\begin{align}
 P_0&=P_{0,r}+\mathcal D_m,\nonumber\\
 P_{0,r}&=f_2\partial_x^2+f_1\partial_x+f_0,\nonumber\\
 P_{1,r}=P_1&=g_1\partial_x+g_0,\nonumber\\
 P_2&=\mathcal H=h_0+a^2\cos^2\theta,\nonumber\\
 \mathcal D_m&=\frac1{\sin\theta}\partial_\theta
                  (\sin\theta\partial_\theta)-m^2\csc^2\theta,
 \label{eq:scalar-polynomial-operators}\\
 f_2&=x^2d,\nonumber\\
 f_1&=2x-\frac{6x^2}{R}+\frac{4cx^3}{R^2}
                       -\frac{2ima x^2}{R},\nonumber\\
 f_0&=-\frac{2x}{R}+\frac{2cx^2}{R^2}-\frac{2ima x}{R},\nonumber\\
 g_1&=2i\left[-R+\frac{\beta x^2}{R}-\frac{4cx^3}{R^2}\right],\nonumber\\
 g_0&=i\left[\frac{2\beta x}{R}-\frac{12cx^2}{R^2}\right]
                  -2ma\left(1+\frac{4x}{R}\right),\nonumber\\
 h_0&=16-4Q^2-a^2\nonumber\\
    &\quad+\frac{(32-8a^2-16Q^2)x}{R}-\frac{16cx^2}{R^2}.
 \label{eq:scalar-radial-coefficients}
\end{align}
The radial coefficients have regular endpoint limits. We form the collocation matrices on $N$ Lobatto nodes and apply inverse $\mathcal H$ there before interpolation. Eliminating the angular dependence gives the radial frequency equation and its total derivative,
\begin{align}
 P_{\rm br}(z)&=P_{0,r}+zP_{1,r}+z^2h_0-A_{11}(az)I,\nonumber\\
 (P_{\rm br})'(z)&=P_{1,r}+2zh_0
                    -\frac{\dd A_{11}(az)}{\dd z}I.
 \label{eq:scalar-branch-pencil}
\end{align}
The frequency derivative of the angular eigenvalue must be included near a double root of this branch.

\subsection{Compactified stress-energy density}
\label{app:scalar-energy-density}

Use $\mu=\cos\theta$ to distinguish the polar variable from frequency $z$, and write $v=i\partial_Tu$, $W=u+x\partial_xu$. Define
\begin{equation}
 \mathcal C_r^{\rm met}=-r^2+\beta-\frac{4c}{r},\qquad
 \mathcal H=h_0+a^2\mu^2.
\end{equation}
With $n_a\dd\Sigma_T=-\delta_a^T\sqrt{-g}\,\dd r\dd\theta\dd\varphi$ and the azimuthal normalization in Eq.~\eqref{eq:scalar-coordinates}, the current energy becomes
\begin{equation}
 E_X=\frac1{2R}\int_0^1\!\dd x\int_{-1}^{1}\!\dd\mu\,
 \mathcal E_X,
 \label{eq:scalar-compact-energy}
\end{equation}
\begin{align}
 \mathcal E_X={}&\mathcal H|v|^2+d|W|^2+(1-\mu^2)|u_\mu|^2
                    +\frac{m^2|u|^2}{1-\mu^2}\nonumber\\
 &-2m\Omega_Z\mathcal H\Rew(\bar vu)\nonumber\\
 &+2m^2\Omega_Za(1+4x/R)|u|^2\nonumber\\
 &+\frac{2mx}{R}(a-\Omega_Z\mathcal C_r^{\rm met})
                          \Imw(\overline Wu).
 \label{eq:scalar-energy-density}
\end{align}
The coefficients are finite at $x=0$, while $a-\Omega_Z\mathcal C_r^{\rm met}=0$ at the horizon. We use this expanded density for quadrature. Its positive, completed-square form is derived in Appendix~\ref{app:hypkn}. Substitution of the state and residual reconstructions below gives the corresponding energy matrices.

The rational coefficient $\Omega_Z$ requires convergence checks on the energy quadrature. The asymptotic weight is also important. For nonzero outgoing data, the energy associated with the unit normal to these asymptotically null slices is generically logarithmically divergent. The current used here gives a finite energy.

\subsection{Angular approximation and finite residual maps}
\label{app:scalar-maps}

At fixed $z$, let $S$ denote the exact continued $S_{11}$ branch and put $f=zu-v$, $h=P_{\rm br}(z)u-P_{1,r}f$. The state and residual are reconstructed at the radial nodes by
\begin{align}
 T_y(u,f)&=\binom{uS}{(zu-f)S},\nonumber\\
 T_r(f,h)&=\binom{fS}{hS/\mathcal H-zfS}.
 \label{eq:scalar-pullback-maps}
\end{align}
We evaluate the second component of $T_r$ at the nodes, interpolate it, and then integrate the energy. This ordering defines the finite calculation.

We compute the angular eigenvalue $A_{11}$ with a truncated continued fraction. The energy integrals use an independently constructed, normalized angular function $S_{K_E}$ from a finite expansion in odd spherical harmonics. Both approximate the same $S_{11}$ branch, but the latter is not obtained by truncating the continued-fraction eigenvector. Table~\ref{tab:scalar-method-settings} lists the two cutoffs. Writing $q_y=(u,f)^T$ and $q_r=(f,h)^T$, the energy matrices in these reduced variables are defined by
\begin{align}
 \|T_y^{(K_E)}q_y\|_{E_X,\rm quad}^2&=q_y^\dagger E_yq_y,
 &E_y&=C_y^\dagger C_y,\nonumber\\
 \|T_r^{(K_E)}q_r\|_{E_X,\rm quad}^2&=q_r^\dagger K_rq_r,
 &K_r&=C_r^\dagger C_r.
 \label{eq:scalar-pullback-grams}
\end{align}
The factors $C_y$ and $C_r$ express these energies as Euclidean norms. They give the finite residual in Eq.~\eqref{eq:constrained}, which approximates the exact angular restriction in Eq.~\eqref{eq:scalar-parent-residual}. Pointwise inversion of $\mathcal H$ and angular truncation do not generally commute at finite resolution. To see the resulting correction, define
\begin{equation}
 e_{K_E}=\mathcal D_mS_{K_E}
       +[a^2z^2\cos^2\theta+A_{11}(az)]S_{K_E},
\end{equation}
Using $S_{K_E}$ in the exact angular residual adds the nodal correction $(0,u e_{K_E}/\mathcal H)^T$ to $T_r^{(K_E)}D_1(u,f)$. This term is omitted by the finite approximation and vanishes for the exact angular function. Refinement of the angular eigenvalue, eigenfunction and energy integrals recovers the exact angular restriction.

Table~\ref{tab:scalar-method-settings} distinguishes the radial resolution, angular-eigenvalue cutoff and energy-integration settings.

\begin{table}[t]
\centering\small
\begin{tabular}{p{0.43\columnwidth}p{0.49\columnwidth}}
\hline
Quantity & Meaning and value \\
\hline
$N$ & $200$ radial nodes (degree $199$) \\
$N_{\rm CF}^{\rm ang}$ & Odd recurrence levels used: $32$ for baseline roots and $40$ for contours and near-EP roots, through $\ell=79$ \\
$K_E$ & $6$ energy harmonics: $\ell=1,3,5,7,9,11$ \\
Energy quadrature & $414\times64$ radial--polar nodes \\
Reduced state dimension & $2N=400$ state/residual coordinates \\
\hline
\end{tabular}
\caption{Scalar radial resolution, angular cutoffs and energy quadrature. Each controls a different approximation.}
\label{tab:scalar-method-settings}
\end{table}

\subsection{Resolution, energy quadrature and working precision}

At the resolutions in Table~\ref{tab:scalar-method-settings}, we compute the radial coefficients, differentiation matrices, roots, mode vectors and frequency derivatives with $200$ decimal working digits. The energy quadrature and Cholesky factors are accurate to 15 to 17 digits. New contour samples use $260$-digit core arithmetic with norm-iteration tolerance $10^{-14}$. Previously computed samples use $200$ digits and tolerance $10^{-8}$. These settings control different errors: working precision does not determine the accuracy of the energy factors or the spatial approximation.

Increasing the energy angular cutoff from six to eight changes the upper factors by less than $6\times10^{-16}$ in the tested samples. With quadrature increased to $600\times120$, the change is at most about $1.4\times10^{-12}$. Comparing the two energy coordinate systems at four frequencies bounds the residual ratios for the same radial matrix within approximately $1\pm1.5\times10^{-10}$.

We also test arithmetic at six frequencies, two per map, while holding the energy factors fixed. The comparison uses $200$-digit core solves at norm-iteration tolerances $10^{-8}$ and $10^{-14}$, together with $260$-digit solves at $10^{-14}$. All eighteen norm iterations converge. Tightening the original settings changes the values by at most $1.32\times10^{-26}$ in relative terms. At the tighter tolerance, changing only the working precision gives a maximum change of $1.42\times10^{-158}$. This test isolates arithmetic error without changing the energy factors.

\subsection{Frequency and double-root checks}

Tables~\ref{tab:revisionroots} and \ref{tab:revision_accuracy} report the baseline roots and comparisons with an independent radial continued-fraction calculation. The reference depths are $8000$ for $n\le5$ and $32000$ for $n\ge6$. At $n=8$, the $N=200$ discrepancies are approximately $4.25\times10^{-5}$ and $6.44\times10^{-5}$ on the positive- and negative-real branches. Increasing to $N=240$ reduces them to $1.07\times10^{-5}$ and $2.35\times10^{-5}$. The much smaller changes with CF depth indicate that radial truncation dominates the error at the deepest poles.

The critical candidate in Eq.~\eqref{eq:scalar-ep-reference} satisfies $F=\partial_\omega F=0$ for the analytic scalar frequency equation. We evaluate this equation with an auxiliary border at $N=200$ and angular cutoff $40$. In the stated border normalization, re-evaluation gives $|F|\simeq1.10\times10^{-184}$, $|\partial_\omega F|\simeq2.85\times10^{-182}$ and $\partial_\omega^2F\simeq(1.21044+0.965555i)\times10^7\ne0$. These residuals support a double root of this finite problem.

At the background in Eq.~\eqref{eq:scalar-enhanced-background}, all four selected roots can be matched one-to-one at radial CF depths $8000,16000,32000,64000$ (Table~\ref{tab:scalar-enhanced-cf-accuracy}). The pair discrepancy is about $4.07\times10^{-32}$, or $4.86\times10^{-23}$ of its gap. Across all four roots, the largest discrepancy is $2.35\times10^{-30}$, and the largest change at the final depth is $2.16\times10^{-46}$. Because the calculations share the angular CF implementation and cutoff, this comparison independently tests only the radial frequencies. It does not test angular truncation or the energy response between poles.

An equivalent scalar double-root condition uses a nullvector $r_0$ and an auxiliary vector $r_1$:
\begin{equation}
 \begin{aligned}
 P_{\rm br}(z_0)r_0&=0,\\
 P_{\rm br}(z_0)r_1+P'_{\rm br}(z_0)r_0&=0.
 \end{aligned}
 \label{eq:hypkn-pencil-Jordan-chain}
\end{equation}
If there is only one independent nullvector and $w$ satisfies $w^\dagger P_{\rm br}(z_0)=0$,
\begin{equation}
 w^\dagger\!\left[P'_{\rm br}(z_0)r_1
       +\tfrac12P''_{\rm br}(z_0)r_0\right]\ne0
 \label{eq:scalar-double-root-order}
\end{equation}
rules out a further vector in the sequence. This condition distinguishes a second-order coalescence from a higher-order degeneracy of the finite frequency equation.

\subsection{Reduced radial solve and spectral instability}

In the coordinates of Eq.~\eqref{eq:scalar-pullback-maps}, the finite weighted residual is $C_rD_1C_y^{-1}$. Its inverse is
\begin{equation}
 D_1(z)^{-1}=\begin{pmatrix}
 P(z)^{-1}P_{1,r}&P(z)^{-1}\\ I&0
 \end{pmatrix},\qquad P(z)=P_{\rm br}(z).
 \label{eq:scalar-inverse-compression}
\end{equation}
This representation preserves both energies and requires high-precision factorization only of the radial matrix. In energy coordinates, the inverse is $C_yD_1^{-1}C_r^{-1}$. A small perturbation $\delta P(z)$ of the radial equation shifts a simple root by
\begin{equation}
 \delta z=-\frac{w^\dagger\delta P(z)r}
                   {w^\dagger P'(z)r},\qquad P(z)=P_{\rm br}(z),
 \label{eq:hypkn-analytic-pencil-shift}
\end{equation}
to first order. The derivative includes the frequency dependence of the angular eigenvalue. At a simple pole,
\begin{equation}
 P(z)^{-1}=\frac{r_jl_j^\dagger}
 {(z-\omega_j)l_j^\dagger P'(\omega_j)r_j}+O(1),
\end{equation}
Taking the norm of this rank-one residue gives Eq.~\eqref{eq:branchkappa}. Local smoothness of the frequency-dependent metrics allows their values at the pole to determine the leading coefficient.

At the tested generic frequencies, the compressed map agrees with the earlier rectangular evaluation to relative error below $10^{-12}$. The scalar coefficient in Eq.~\eqref{eq:branchkappa} applies to the restricted angular branch. It has a different scope from the full-operator condition number in Eq.~\eqref{eq:paper-kappa}.

The measured local slopes satisfy $\sigma_1^{G_N}\kappa_{1,E_X}/|z-\omega_j|=1$ within $2.5\times10^{-11}$ for the positive baseline fundamental. For near-EP 1 at real displacement $|\delta M\omega|=10^{-10}d$, the ratio is $0.9999999999212393$. These tests verify the pole normalization at fixed resolution. The main text discusses how the condition numbers depend on radial resolution.

\subsection{Linear solves and frequency-plane interpolation}

The baseline, wide EP and local EP maps use $\ScalarBaseSamples$, $\ScalarWideSamples$ and $\ScalarLocalSamples$ sampled frequencies, respectively. Previously computed $200$-digit samples contribute $2148$, $1036$, $377$ values, and new $260$-digit samples contribute $763$, $1366$, $870$. The wide EP map also uses the local samples. The $2177$ distinct new evaluations comprise $2129$ map values and $48$ fixed checks, with sixteen checks per view. Every new solve meets the norm-iteration tolerance $10^{-14}$. Each sample is evaluated at its own background, with the corresponding radial matrix and energy norm.

We construct the contours and color fields by piecewise-linear Delaunay interpolation of measured $\log_{10}\sigma$, excluding the check frequencies from the fit. The maximum discrepancies at those checks are $0.03343$, $0.04204$, $0.002284$ decades for the baseline, wide EP and local EP maps. These sampled discrepancies do not establish a uniform error bound. Rounded line joins change only the rendering.

The radial solve uses an equivalent banded Chebyshev--ultraspherical representation. If $V_N$ maps coefficients to nodes and $\mathcal U_N$ converts to the second ultraspherical basis, the internal matrix is
\begin{align}
 \mathcal B_N&=\mathcal U_NV_N^{-1}P_r(z)V_N,\nonumber\\
 P_r(z)^{-1}&=V_N\mathcal B_N^{-1}\mathcal U_NV_N^{-1}.
\end{align}
Polynomial multiplication must respect the Lobatto alias identities to preserve the nodal equation. With $n=N-1$ and $\xi=2x-1$, these are
\begin{equation}
 \begin{aligned}
 T_{n+j}(\xi_k)&=T_{n-j}(\xi_k),\\
 \xi_k&=\cos(k\pi/n),\qquad 0\leq j,k\leq n.
 \end{aligned}
 \label{eq:hypkn-Lobatto-alias}
\end{equation}
Dropping coefficients above degree $n$ would instead change the finite matrix. At $N=200$, a direct test of nodal equivalence gives a difference of about $8\times10^{-193}$. The tested physical residuals agree with dense solves through all $50$ displayed digits.

These checks address distinct sources of error in the finite calculation: frequencies, energy factors, arithmetic and interpolation. Continuum convergence and the outgoing-wave domain require separate consideration, as discussed in Sec.~\ref{sec:discussion} and Appendix~\ref{sec:domain}.

\begin{table*}[t]
\centering\small
\begin{tabular}{r r r r r}
\hline
$n$ & $\Rew\omega_+$ & $\Imw\omega_+$ & $\Rew\omega_-$ & $\Imw\omega_-$ \\
\hline
0 & 0.3313102486 & -0.0970717672 & -0.2794451001 & -0.0980515141 \\
1 & 0.3086464999 & -0.3007691861 & -0.2483426962 & -0.3093138466 \\
2 & 0.2784702478 & -0.5234694229 & -0.2119598677 & -0.5487579540 \\
3 & 0.2527283824 & -0.7590255324 & -0.1864852668 & -0.8023696778 \\
4 & 0.2328644684 & -0.9998314052 & -0.1699096919 & -1.0591718009 \\
5 & 0.2171693493 & -1.2424475566 & -0.1584714332 & -1.3160558842 \\
6 & 0.2041820994 & -1.4856308998 & -0.1500085265 & -1.5724005239 \\
7 & 0.1929753620 & -1.7289287450 & -0.1433820698 & -1.8281632374 \\
8 & 0.1829597680 & -1.9721135486 & -0.1380383294 & -2.0834604829 \\
\hline
\end{tabular}
\caption{Selected neutral massless scalar $\ell=m=1$ roots at $(a,Q)=(0.3,0.4)$, with $N=200$ radial nodes and $200$ decimal working digits. The displayed digits identify the finite-grid roots. Table~\ref{tab:revision_accuracy} compares them with continuum-frequency references. Overtone labels are continued independently on the two real-frequency branches.}
\label{tab:revisionroots}
\end{table*}

\begin{table}[t]
\centering\small
\begin{tabular}{rcc}
\hline
$n$ & $|M\omega_{+,200}-M\omega_{+,\mathrm{CF}}|$ & $|M\omega_{-,200}-M\omega_{-,\mathrm{CF}}|$ \\
\hline
0 & $8.78715\times10^{-36}$ & $6.18410\times10^{-33}$ \\
1 & $2.78072\times10^{-24}$ & $1.71303\times10^{-20}$ \\
2 & $1.02637\times10^{-16}$ & $3.51805\times10^{-13}$ \\
3 & $2.83074\times10^{-12}$ & $1.49033\times10^{-9}$ \\
4 & $1.30875\times10^{-9}$ & $1.21407\times10^{-7}$ \\
5 & $6.98651\times10^{-8}$ & $1.63176\times10^{-6}$ \\
6 & $1.11928\times10^{-6}$ & $8.76423\times10^{-6}$ \\
7 & $8.69631\times10^{-6}$ & $2.79707\times10^{-5}$ \\
8 & $4.24579\times10^{-5}$ & $6.44476\times10^{-5}$ \\
\hline
\end{tabular}
\caption{Absolute frequency differences between the $N=200$ scalar roots and independent Leaver continued-fraction references at $(a/M,Q/M)=(0.3,0.4)$, $\ell=m=1$. The two frequency-sign branches are matched separately. Both calculations use $200$ working digits. The errors chiefly reflect radial discretization.}
\label{tab:revision_accuracy}
\end{table}

\begin{table}[t]
\centering\small
\begin{tabular}{lcc}
\hline
Mode & $\Delta_{\rm CF}$ & $\Delta_{\rm depth}$ \\
\hline
near-EP 1 & $4.0675\times10^{-32}$ & $2.1640\times10^{-46}$ \\
near-EP 2 & $4.0675\times10^{-32}$ & $2.1640\times10^{-46}$ \\
QNM 1 & $3.9935\times10^{-34}$ & $2.7866\times10^{-52}$ \\
QNM 2 & $2.3484\times10^{-30}$ & $2.8017\times10^{-51}$ \\
\hline
\end{tabular}
\caption{Radial continued-fraction comparison for the four scalar roots at Eq.~\eqref{eq:scalar-enhanced-background}, with one-to-one matching and $200$ working digits. Here $\Delta_{\rm CF}=|M\omega_{200}-M\omega_{\mathrm{CF},64000}|$ and $\Delta_{\rm depth}=|M\omega_{\mathrm{CF},64000}-M\omega_{\mathrm{CF},32000}|$. The shared angular calculation makes this a radial-frequency check.}
\label{tab:scalar-enhanced-cf-accuracy}
\end{table}
\section{Coupled GE equations and radiative energy}
\label{app:ge-energy}
\label{app:ge-symmetrizer}
\label{app:ge}

This appendix derives the positive radiative energy used in Section~\ref{sec:ge} from the coupled field equations. We use the coordinates, background functions and endpoint locations of Appendix~\ref{app:hypkn}.

\subsection{Covariant coupled Einstein--Maxwell system}
\label{app:hypkn-GE-covariant}

In our normalization, the background Einstein--Maxwell equations read
\begin{align}
 R_{ab}&=2\left(F_{ac}F_b{}^c-\frac14g_{ab}F_{cd}F^{cd}\right),
 \nonumber\\
 \nabla_aF^{ab}&=0,\qquad \nabla_{[a}F_{bc]}=0.
 \label{eq:hypkn-background-EM}
\end{align}
The background potential $A=-Qr\Sigma^{-1}(dv-a\sin^2\theta\,d\varphi)$ is regular at the horizon. In compact coordinates its radial component develops a pure-gauge pole at infinity. Subtracting $2Q\,d\ln x$ gives
\begin{align}
 A'&=A'_T\,dT+A'_x\,dx+A'_\varphi\,d\varphi,
 \nonumber\\
 S_0&=R^2+a^2x^2\cos^2\theta,
 \nonumber\\
 A'_T&=-\frac{QRx}{S_0},
 \nonumber\\
 A'_x&=\frac{4QR-2Qa^2x\cos^2\theta}{S_0},
 \nonumber\\
 A'_\varphi&=\frac{QaRx\sin^2\theta}{S_0}.
 \label{eq:hypkn-compact-potential}
\end{align}
The potential components and the unchanged field strength $F=dA'=dA$ are regular on the compact exterior.

Write the metric and electromagnetic perturbations as $h_{ab}=\delta g_{ab}$ and $b_a=\delta A_a$, with $f_{ab}=2\nabla_{[a}b_{b]}$. Here $b_a$ denotes the perturbed potential and is unrelated to the coordinate function $b(r)$. We raise perturbation indices with the background metric and write $h=g^{ab}h_{ab}$. The required variations are
\begin{align}
 \delta\Gamma^c{}_{ab}
 &=\frac12g^{cd}\bigl(\nabla_ah_{bd}+\nabla_bh_{ad}
 \nonumber\\
 &\hspace{4.7em}-\nabla_dh_{ab}\bigr),
 \nonumber\\
 \delta R_{ab}
 &=\nabla_c\delta\Gamma^c{}_{ab}
                  -\nabla_b\delta\Gamma^c{}_{ac},
 \nonumber\\
 \delta(F_{cd}F^{cd})
 &=2F^{cd}f_{cd}-2h^{ce}F_c{}^dF_{ed},
 \nonumber\\
 \delta S_{ab}
 &=f_{ac}F_b{}^c+F_{ac}f_b{}^c
 \nonumber\\
 &\quad-h^{cd}F_{ac}F_{bd}-\frac14h_{ab}F_{cd}F^{cd}
 \nonumber\\
 &\quad-\frac14g_{ab}\,\delta(F_{cd}F^{cd}).
 \label{eq:hypkn-linearized-EM-definitions}
\end{align}
The variation of the Maxwell density is conveniently written in terms of
\begin{align}
 \mathcal M^{ab}&=f^{ab}-h^a{}_cF^{cb}-h^b{}_cF^{ac}
 \nonumber\\
 &\quad+\frac12hF^{ab}.
 \label{eq:hypkn-Maxwell-density-amplitude}
\end{align}
The exact linearized coupled equations are
\begin{align}
 \delta R_{ab}-2\delta S_{ab}&=0,
 \nonumber\\
 \nabla_a\mathcal M^{ab}&=0,\qquad
 \nabla_{[a}f_{bc]}=0.
 \label{eq:hypkn-linearized-covariant-system}
\end{align}
The $hF^{ab}/2$ term includes the variation of the volume element in $\partial_a(\sqrt{-g}F^{ab})=0$.

For a potential formulation one may impose background de Donder and Lorenz gauges,
\begin{align}
 \nabla^a\left(h_{ab}-\frac12g_{ab}h\right)&=0,
 \nonumber\\
 \nabla^ab_a&=0.
 \label{eq:hypkn-GE-gauges}
\end{align}
The highest-derivative terms are wave operators on $h_{ab}$ and $b_a$. Couplings through the background $F$ enter at lower derivative order, so the wavefronts follow the background light cones. Physical GE QNMs must also satisfy the constraints and be distinguished from residual gauge solutions. A coordinate change generated by $\xi$ and an electromagnetic gauge change $\chi_{\rm g}$ act as $h\mapsto h+\mathcal L_\xi g$ and $b\mapsto b+\mathcal L_\xi A'+d\chi_{\rm g}$.

The radiative variables introduced below describe this coupled metric--Maxwell system.

\subsection{Regular four-field GE radiative system}
\label{app:hypkn-GE-radiative}

For $Q\ne0$, we use the coupled positive-spin equations of Ref.~\cite{Hintz2026} in their four-field, first-order form. The following definitions fix the field normalization.

Set $\zeta=r+ia\cos\theta$, $\chi=r-ia\cos\theta$, and introduce the rescaled Newman--Penrose amplitudes
\begin{align}
 \psi_2&=\Psi_0^{(1)},\qquad
 \psi_1=\sqrt2\,\chi\Psi_1^{(1)},
 \nonumber\\
 k_{\rm NP}&=\frac{\kappa^{(1)}}{\sqrt2\,\chi^2},
 \nonumber\\
 s_{\rm NP}&=\frac{\zeta\sigma^{(1)}}{\chi^2}.
 \label{eq:hypkn-GE-positive-fields}
\end{align}
The connection amplitude $k_{\rm NP}$ is unrelated to the azimuthal function $k(r)$. Likewise, the Newman--Penrose spin coefficient $\kappa^{(1)}$ should not be confused with surface gravity or a mode condition number. When the background Coulomb field is nonzero, the phantom tetrad gauge sets the extreme Maxwell perturbations to zero through a tetrad choice. Electromagnetic radiation is still present in the coupled Weyl and connection amplitudes. We follow the Newman--Penrose normalization of Ref.~\cite{Hintz2026} throughout. Any change of curvature-sign convention or tetrad must be applied consistently to both the state and its norm.

For a Boyer--Lindquist harmonic factor $e^{-izt+im\phi}$, define
\begin{align}
 K_r&=am-A_rz,
 \nonumber\\
 K_\theta&=m\csc\theta-az\sin\theta,
 \nonumber\\
 \mathcal D_j&=\partial_r+\frac{iK_r}{\Delta}
                                   +\frac{j\Delta'}{\Delta},
 \nonumber\\
 \mathcal D_j^\dagger&=\partial_r-\frac{iK_r}{\Delta}
                                   +\frac{j\Delta'}{\Delta},
 \nonumber\\
 \mathcal L_j&=\partial_\theta+K_\theta+j\cot\theta,
 \nonumber\\
 \mathcal L_j^\dagger&=\partial_\theta-K_\theta+j\cot\theta,
 \nonumber\\
 F_\pm&=3\left(1-\frac{Q^2}{\zeta}\right)
                  \pm\frac{Q^2\chi}{\zeta^2},
 \nonumber\\
 Z&=\frac{\zeta}{\chi^2}.
 \label{eq:hypkn-GE-building-blocks}
\end{align}
The dagger denotes the sign changes shown above, without conjugating the complex frequency. Introduce
\begin{align}
 \mathcal U&=\mathcal L_2-\frac{3ia\sin\theta}{\chi},
 \nonumber\\
 \mathcal R&=\mathcal D_0+\frac3\chi,
 \nonumber\\
 \mathcal V&=\mathcal L_{-1}^\dagger+\frac{3ia\sin\theta}{\chi},
 \nonumber\\
 \mathcal T&=\Delta\left(\mathcal D_2^\dagger-\frac3\chi\right).
 \label{eq:hypkn-GE-first-order-factors}
\end{align}
The original positive-spin system is then
\begin{align}
 \mathcal U\psi_2-\mathcal R\psi_1&=-2F_+k_{\rm NP},
 \nonumber\\
 \mathcal T\psi_2+\mathcal V\psi_1&=2F_-s_{\rm NP},
 \nonumber\\
 \mathcal R s_{\rm NP}-\mathcal V k_{\rm NP}&=Z\psi_2,
 \nonumber\\
 \mathcal T k_{\rm NP}+\mathcal U s_{\rm NP}&=2Z\psi_1.
 \label{eq:hypkn-GE-original-first-order}
\end{align}
Derivatives act on all coefficients and fields to their right.

\subsection{GE rescaling and endpoint regularity}
\label{app:hypkn-GE-regularization}

The fields $(\psi_2,\psi_1,k_{\rm NP},s_{\rm NP})$ have boost weights $(2,1,2,1)$ and spin weights $(2,1,1,2)$. A field of boost weight $b$ has physical oscillatory asymptotics
\begin{align}
 w_b^{\rm BL}&\sim\Delta^{-b}e^{-i(z-m\Omega_H)r_*}
 \quad(r\to R),
 \nonumber\\
 w_b^{\rm BL}&\sim r^{-1-2b}e^{+izr_*}
 \quad(r\to\infty).
 \label{eq:hypkn-GE-asymptotics}
\end{align}
Both endpoint behaviors are factored out by writing
\begin{equation}
 w_b^{\rm BL}=e^{-iz\mathfrak h(r)+imk(r)}
                        \frac{\Delta^{-b}}{r}\,u_b(x,\theta).
 \label{eq:hypkn-GE-individual-rescaling}
\end{equation}
This gives the regular state $Y$ of Eq.~\eqref{eq:ge-regular-state} in $(T,r,\theta,\varphi)$, with the common azimuthal factor suppressed. The factor $2Q$ sets the relative normalization of the curvature and connection fields. This rescaling is invertible at all charged backgrounds considered here. Taking $Q\to0$ would require a representation regular in that limit.

Substituting Eq.~\eqref{eq:hypkn-GE-individual-rescaling} gives the radial operators acting on a field of boost weight $b$
\begin{align}
 d_{j,b}^\dagger&=-\frac{x^2}{R}\partial_x
 +(j-b)\frac{\Delta'}{\Delta}-\frac1r
 \nonumber\\
 &\quad+iz\left(2+\frac4r\right),
 \nonumber\\
 d_{j,b}&=d_{j,b}^\dagger+\frac{2iK_r}{\Delta}.
 \label{eq:hypkn-GE-transformed-radial}
\end{align}
The boost weight, rather than the spin weight, enters this radial substitution.

Eliminating two fields would require division by $F_-$. At the equator,
\begin{equation}
 F_-\big|_{\theta=\pi/2}=3-\frac{4Q^2}{r}.
 \label{eq:hypkn-GE-alpha-pole}
\end{equation}
The zero $r_p=4Q^2/3$ lies outside the horizon at the highly charged near-EP background and inside it at $(a,Q)=(0.3,0.4)$~\cite{DiasGodazgarSantos2015,Hintz2026}. This singularity arises from the two-field reduction. The four-field equations involve no division by $F_-$ and remain regular there.

\subsection{Transformation of the principal matrices}

For the state $U$ of Eq.~\eqref{eq:ge-base-state}, we place all terms on the left and order the equations as $(-1b,-1a,-1d,+1c)$ in Ref.~\cite{Hintz2026}. In Boyer--Lindquist coordinates this gives
\begin{equation}
 A^t\partial_tU+A^r\partial_rU+A^\theta\partial_\theta U
              +A^\phi\partial_\phi U+C_{\rm BL}U=0.
 \label{eq:ge-BL-system}
\end{equation}
Its two principal blocks, labelled by $\eta=\pm1$, are
\begin{align}
 A^t_\eta&=\begin{pmatrix}
 A_r&-i\eta a\sin\theta\\i\eta a\sin\theta&A_r/\Delta
 \end{pmatrix},\nonumber\\
 A^r_\eta&=\begin{pmatrix}-\Delta&0\\0&1\end{pmatrix},
 \qquad A^\theta_\eta=-\begin{pmatrix}0&1\\1&0\end{pmatrix},\nonumber\\
 A^\phi_\eta&=\begin{pmatrix}
 a&-i\eta\csc\theta\\i\eta\csc\theta&a/\Delta
 \end{pmatrix}.
 \label{eq:ge-BL-principal}
\end{align}
These matrices are Hermitian, with $\det A^t_\eta=\mathcal A/\Delta>0$ in the open exterior. Apply the coordinate change of Eq.~\eqref{eq:hypkn-coordinates} and write $U=FY$, where
\begin{equation}
 F=\operatorname{diag}\bigl((r\Delta^2)^{-1},(r\Delta)^{-1},
                         (r\Delta^2)^{-1},(r\Delta)^{-1}\bigr),
 \label{eq:ge-radial-field-map}
\end{equation}
before multiplying on the left by $r^2\Delta^3F^\dagger$ to obtain Eq.~\eqref{eq:ge-hyp-system}. The identity $2A_r-(2+4/r)\Delta=8-2Q^2-4c_0/r$ gives the finite lower-right time coefficient in Eq.~\eqref{eq:ge-Bblock}.

\subsection{Regular coefficient matrices}
\label{app:hypkn-GE-coefficients}

In the stated state ordering the fixed-$m$ evolution equations are
\begin{equation}
 B\partial_TY+\mathsf A^x\partial_xY+\mathsf A^\theta\partial_\theta Y+\mathsf C_mY=0.
 \label{eq:hypkn-GE-first-order-form}
\end{equation}
The time matrix is $B=\operatorname{diag}(B_+,B_-)$, with the blocks in Eq.~\eqref{eq:ge-Bblock}. Write $Y=(Y_+,Y_-)^T$ and use $b=2+4x/R$, $d=8-2Q^2-4c_0x/R$ below. The spatial derivative matrices are
\begin{align}
 \mathsf A^x&=\operatorname{diag}(x^2/R,-Rf,x^2/R,-Rf),
 \nonumber\\
 \mathsf A^\theta&=\begin{pmatrix}
 0&-1&0&0\\-1&0&0&0\\0&0&0&-1\\0&0&-1&0
 \end{pmatrix}.
 \label{eq:hypkn-GE-spatial-matrices}
\end{align}
Here $f(x)$ is defined in Eq.~\eqref{eq:hypkn-geometric-coefficients}. All radial coefficients have finite limits at both endpoints.

For a lower-order matrix free of large-$r$ cancellation, set $\nu=x/R$, $\chi_0=1-ia\nu\cos\theta$ and $\zeta_0=1+ia\nu\cos\theta$. Define
\begin{align}
 c_1&=\nu+\frac{3\nu}{\chi_0},
 \nonumber\\
 c_2&=4-c_0\nu+2iam
 \nonumber\\
 &\quad+\frac{-6+3c_0\nu+3ia\cos\theta}{\chi_0},
 \nonumber\\
 a_m&=\frac{m+\cos\theta}{\sin\theta}
                       -\frac{3ia\nu\sin\theta}{\chi_0},
 \nonumber\\
 b_m&=-\frac{m+2\cos\theta}{\sin\theta}
                       +\frac{3ia\nu\sin\theta}{\chi_0},
 \nonumber\\
 F_\pm&=3\left(1-\frac{Q^2\nu}{\zeta_0}\right)
                \pm\frac{Q^2\nu\chi_0}{\zeta_0^2},
 \nonumber\\
 Z&=\frac{\nu\zeta_0}{\chi_0^2}.
 \label{eq:hypkn-GE-regular-potentials}
\end{align}
The complete potential is
\begin{equation}
 \mathsf C_m=\begin{pmatrix}
 c_1&a_m&0&F_-/Q\\
 b_m&c_2&-F_+/Q&0\\
 0&4QZ&c_1&b_m\\
 -2QZ&0&a_m&c_2
 \end{pmatrix}.
 \label{eq:hypkn-GE-potential-matrix}
\end{equation}
In the original radial coordinate, the second diagonal coefficient is $c_2=-\Delta/r-\Delta'+3\Delta/\chi+2iam$. The rational form above displays its finite limit directly. Neither $\chi_0$ nor $\zeta_0$ vanishes on the real exterior. The apparent polar singularities in $a_m,b_m$ combine with angular derivatives when acting on spin-regular fields.

Before azimuthal decomposition one has $\mathsf C_m=\mathsf C_{0}+im\mathsf A^\varphi$, where
\begin{align}
 \mathsf A^\varphi&=\operatorname{diag}(\mathsf A_+^\varphi,\mathsf A_-^\varphi),
 \nonumber\\
 \mathsf A_\eta^\varphi&=\begin{pmatrix}
 0&-i\eta\csc\theta\\i\eta\csc\theta&2a
 \end{pmatrix},\qquad \eta=\pm1.
 \label{eq:hypkn-GE-azimuth-matrix}
\end{align}
Hermiticity also holds for this matrix, and hence for all derivative matrices before azimuthal decomposition. For a real wave covector $\xi_\mu$, the determinant of each two-component derivative block is $-\Sigma g^{\mu\nu}\xi_\mu\xi_\nu$. The full determinant is its square, so the regular radiative equations propagate on the background light cones.

For $Y=e^{-izT}y$, the frequency-domain equation is
\begin{equation}
 P(z)y=(\mathcal D_m-izB)y=0,
 \label{eq:hypkn-GE-pencil-generator}
\end{equation}
where $\mathcal D_m$ and the evolution operator are defined in Eq.~\eqref{eq:ge-generator}. No frequency-dependent angular separation has been used.

\subsection{Positive energy and balance law}
\label{app:hypkn-GE-energy}

In addition to Hermiticity of the principal matrices, we need positivity of the time coefficient. Its determinant satisfies
\begin{align}
 \det B_\eta&=bd-a^2\sin^2\theta
 \nonumber\\
 &=H(x,\theta)>0,\qquad b\geq2.
 \label{eq:hypkn-GE-positive-determinant}
\end{align}
The exact eigenvalues are
\begin{equation}
 \lambda_\pm(B_\eta)=\frac{b+d\pm
       \sqrt{(b-d)^2+4a^2\sin^2\theta}}2.
 \label{eq:ge-time-eigenvalues}
\end{equation}
Both eigenvalues are positive. Since $d>2$ throughout the subextremal compact exterior, elementary bounds give $\lambda_{\min}(B_\eta)\geq2-|a|>1$ and $\lambda_{\max}(B_\eta)<9$. The quadratic form is uniformly positive for the chosen regularized state, including at the endpoints.

With $d\mu=\sin\theta\,dx\,d\theta\,d\varphi$, the energy in Eq.~\eqref{eq:ge-energy} becomes, in one azimuthal sector,
\begin{align}
 E_{\rm GE}[Y]&=\pi\int_0^1\!dx\int_0^\pi\!d\theta\,\sin\theta\,
                       \mathcal E_{\rm GE}(Y),\nonumber\\
 \mathcal E_{\rm GE}(Y)&=Y_+^\dagger B_+Y_++Y_-^\dagger B_-Y_-.
 \label{eq:hypkn-GE-energy-explicit}
\end{align}
Without the $2Q$ connection rescaling, the same energy carries a factor $4Q^2$ on the connection block. Any field redefinition must be accompanied by the corresponding change of energy matrix. Section~\ref{sec:discussion} discusses the physical interpretation of this radiative norm.

For the fixed-$m$ system with source $j$ and coordinate density $\rho=\sin\theta$, define the production matrix
\begin{equation}
 \mathcal Q_E=\rho^{-1}\partial_i(\rho\mathsf A^i)
                        -\mathsf C_m-\mathsf C_m^\dagger.
 \label{eq:hypkn-GE-production-matrix}
\end{equation}
The index $i$ ranges over $x,\theta$. Direct integration by parts gives
\begin{align}
 \frac{dE_{\rm GE}}{dT}={}&-\frac12\int_{\partial\Sigma_T}
       \rho Y^\dagger\mathsf A^i\nu_iY\,dS
 \nonumber\\
 &+\frac12\int_{\Sigma_T}\rho Y^\dagger\mathcal Q_EY\,d^3x
 \nonumber\\
 &+\operatorname{Re}\int_{\Sigma_T}\rho Y^\dagger j\,d^3x.
 \label{eq:ge-balance}
\end{align}
The potentially singular angular terms cancel. For example, writing $\mathsf C=\mathsf C_{m=0}$,
\begin{equation}
 \mathsf C_{12}+\overline{\mathsf C_{21}}+\cot\theta
 =-\frac{6iar\sin\theta}{r^2+a^2\cos^2\theta}.
 \label{eq:ge-angular-cancellation}
\end{equation}
These cancellations make $\mathcal Q_E$ bounded at each fixed subextremal background. Regular angular fields contribute no boundary terms at the poles. The remaining radial flux is
\begin{equation}
 \mathcal F_{\partial\Sigma}=
 \frac12\int\!\left[Y^\dagger\mathsf A^xY\right]_{x=0}^{x=1}
                    \sin\theta\,d\theta\,d\varphi.
 \label{eq:ge-boundary-flux}
\end{equation}
At $x=0$ and $x=1$, the radial blocks are $\operatorname{diag}(0,-R)$ and $\operatorname{diag}(1/R,0)$, respectively, giving a nonnegative outward flux. The limiting field equations govern the zero-speed components without independent incoming boundary data. For a source-free solution, define
\begin{equation}
 C_E=\sup_{x,\theta}
 \|B^{-1/2}\mathcal Q_E B^{-1/2}\|_2<\infty.
 \label{eq:ge-production-bound}
\end{equation}
The balance law implies $\dd E_{\rm GE}/\dd T\leq C_E E_{\rm GE}$. Positivity of the norm does not imply conservation or monotone decay, because the interior term has no definite sign. Appendix~\ref{sec:domain} addresses the additional endpoint conditions needed for the QNM problem.
\section{Coupled GE discretization and numerical checks}
\label{app:numerics}
\label{app:ge-validation}
\label{app:hypkn-GE-angular}

At fixed $m$, all four radiative fields are expanded with their full angular dependence. We use the radial nodes and polynomial identities of Eqs.~\eqref{eq:hypkn-radial-grid} and \eqref{eq:hypkn-Lobatto-alias}. After removing $e^{im\varphi}$, the spin weights and lowest angular degrees of the components are
\begin{equation}
 \begin{aligned}
 (s_1,s_2,s_3,s_4)&=(2,1,1,2),\\
 \ell_{\min,i}&=\max(|m|,|s_i|).
 \end{aligned}
 \label{eq:ge-angular-truncation-indices}
\end{equation}
For $y=\cos\theta$ and spin weight $s$, define
\begin{equation}
 \begin{aligned}
 \alpha_s&=|m+s|,\qquad \beta_s=|m-s|,\\
 j_\ell&=\ell-\max(|m|,|s|).
 \end{aligned}
 \label{eq:hypkn-GE-angular-indices}
\end{equation}
The regular angular functions are represented by Jacobi polynomials:
\begin{align}
 p_{\ell m}^{(s)}(y)={}&\mathcal N_{\ell m}^{(s)}
 (1-y)^{\alpha_s/2}(1+y)^{\beta_s/2}\nonumber\\
 &\times P_{j_\ell}^{(\alpha_s,\beta_s)}(y),
 \label{eq:hypkn-spin-poles}
\end{align}
with normalization
\begin{equation}
 \int_{-1}^{1}p_{\ell m}^{(s)*}p_{\ell' m}^{(s)}\,\dd y
 =\delta_{\ell\ell'}.
 \label{eq:hypkn-GE-angular-normalization}
\end{equation}
The finite trial state is expanded as
\begin{equation}
 \begin{aligned}
 Y_{i,N,L_{\rm ang}}^{(m)}(T,x,\theta)
 ={}&\sum_{\ell=\ell_{\min,i}}^{L_{\rm ang}}
       \sum_{j=0}^{N-1} y_{i\ell j}(T)\,b_j(x)\\
    &\hspace{1.5em}\times p_{\ell m}^{(s_i)}(\cos\theta).
 \end{aligned}
 \label{eq:ge-angular-truncated-state}
\end{equation}
The radial interpolation polynomials $b_j(x_k)=\delta_{jk}$ have degree $N-1$, and $y_{i\ell j}(T)$ are complex coefficients. The resolution parameters play different roles: $N$ counts radial nodes, whereas $L_{\rm ang}$ sets the largest angular degree in the expansion. Angular quadrature order is chosen separately.

The evolution matrix, frequency-dependent matrix and energy matrix depend on both resolutions and on $m$. We write
\begin{equation}
 L_{N,L_{\rm ang}}^{(m)},\qquad
 P_{N,L_{\rm ang}}^{(m)}(z),\qquad
 G_{N,L_{\rm ang}}^{(m)}.
 \label{eq:ge-resolution-notation}
\end{equation}
With $L_{\rm ang}$ and $m$ fixed, we use the shorter notation $L_N$, $P_N$ and $G_N$.

For $L_{\rm ang}\geq\max(2,|m|)$, the number of complex state coefficients is
\begin{equation}
 d(N,L_{\rm ang},m)
 =N\sum_{i=1}^{4}\bigl(L_{\rm ang}-\ell_{\min,i}+1\bigr).
 \label{eq:ge-truncation-dimension}
\end{equation}
The two GE pseudospectral panels in Fig.~\ref{fig:ge-pair} use $N=200$ and $L_{\rm ang}=6$. Working precision is specified below. At $m=1$, the four components contain $5,6,6,5$ angular functions, respectively, giving $d=4400$. At $m=2$, each contains $5$, giving $d=4000$. The $m=1$ calculation includes the spin-one dipole sector, whereas all four $m=2$ expansions begin at $\ell=2$. A label such as gravitational-led $\ell=2$ identifies a mode by continuation. The operator itself includes every angular family within the chosen truncation.

\subsection{Energy matrices and condition numbers}

Before angular projection, the numerical solver multiplies the equations by the positive, frequency-independent factor
\begin{equation}
 f_c(x,\theta)=\left(1+\frac{a^2x^2\cos^2\theta}{R^2}\right)^2.
 \label{eq:hypkn-clearing-factor}
\end{equation}
This clears rational coordinate denominators while leaving the continuous equations unchanged. At finite $L$, projecting $f_c(\mathcal D_m-izB)$ need not be equivalent to projecting $\mathcal D_m-izB$. We use the former prescription, along with the radial basis conversion and Lobatto identities cited above. These row and basis transformations enter the mass/source factor of Eq.~\eqref{eq:ge-forced-inverse}. The energy is still computed from the original $B$-quadratic form, not from the time matrix after multiplication by $f_c$.

For sampled basis columns $\Phi_N$, $G_N=\tfrac12\int\Phi_N^\dagger B\Phi_N\,\dd\mu$. Since $B$ is affine in $x$, products of degree-$(N-1)$ radial interpolants have energy integrands of degree at most $2N-1$. An $N$-node Gauss--Legendre rule integrates these radial products exactly in exact arithmetic. Angular products are evaluated analytically through Jacobi-polynomial beta-function formulas. We apply $C_N$ and $C_N^{-1}$ using a radial evaluation matrix and small angular Cholesky factors, avoiding dense factorization at the full state dimension.

For a simple eigenvalue of $L_N$, let $v_j$ and $y_j^\dagger$ be its right and left eigenvectors. The condition number of Section~\ref{sec:framework} becomes
\begin{equation}\label{eq:ge-kappa}
 \kappa_E(\omega_j)=
 \frac{\sqrt{v_j^\dagger G_Nv_j}\,
       \sqrt{y_j^\dagger G_N^{-1}y_j}}
      {|y_j^\dagger v_j|}.
\end{equation}
If the eigenproblem is stored as $K_Nv_j=\omega_jM_Nv_j$ and $w_j^\dagger K_N=\omega_jw_j^\dagger M_N$, then $y_j=M_N^\dagger w_j$ and
\begin{equation}\label{eq:ge-gen-kappa}
 \kappa_E(\omega_j)=
 \frac{\sqrt{v_j^\dagger G_Nv_j}\,
 \sqrt{w_j^\dagger M_NG_N^{-1}M_N^\dagger w_j}}
 {|w_j^\dagger M_Nv_j|}.
\end{equation}
Here $K_N=-iD_N$ and $M_N=B_N$. The mass and Gram matrices have distinct roles. If their coefficients are perturbed separately, the frequency shift is
\begin{equation}
 \delta\omega_j=\frac{w_j^\dagger
       (\delta K_N-\omega_j\delta M_N)v_j}
       {w_j^\dagger M_Nv_j}+O(\delta^2),
\end{equation}
for specified coefficient weights \cite{Tisseur2000}. For a frequency-independent, invertible field change $Y'=TY$, the corresponding matrices are
\begin{equation}
 \begin{aligned}
 L_N'&=TL_NT^{-1},\\
 G_N'&=(T^{-1})^\dagger G_NT^{-1}.
 \end{aligned}
 \label{eq:ge-field-energy-change}
\end{equation}
These transformations leave the energy of each physical state and its resolvent norm unchanged.

At an exact second-order EP, the eigenvector $r_0$, its associated vector $r_1$, and a left eigenvector $\ell_0^\dagger$ satisfy
\begin{equation}
 \begin{gathered}
 P_{\rm EP}r_0=0,\qquad P_{\rm EP}r_1=iB_Nr_0,\\
 \ell_0^\dagger P_{\rm EP}=0
       \quad\Longrightarrow\quad\ell_0^\dagger B_Nr_0=0.
 \end{gathered}
\end{equation}
The denominator of Eq.~\eqref{eq:ge-gen-kappa} vanishes while its numerator remains nonzero, accounting for the divergence of the individual-mode condition number on approach to the EP.

\subsection{Arithmetic and independent checks}
\label{app:ge-arithmetic}

The QNM and left/right-vector calculations use $260$ working digits, a $100$-digit Newton-step target and a linear backward residual target of $10^{-200}$. Coefficient assembly, energy factors, residuals and norm iteration for the contour samples also use $260$ digits. The linear solves target a backward residual of $10^{-60}$ and fall back to full arbitrary-precision factorization when needed. A reorthogonalized inverse-norm Lanczos iteration targets a relative Ritz residual of $10^{-10}$. Binary64 sparse LU is used only as a preconditioner. These stopping criteria neither certify hundreds of digits in the singular values nor provide bounds enclosing the global inverse norm.

Near the pair, the inverse is evaluated efficiently by a temporary rank-one update and its exact correction:
\begin{equation}\label{eq:ge-border-inverse}
 \begin{gathered}
 \widetilde P_N=P_N+e_0e_0^\dagger,\qquad
 u_N=\widetilde P_N^{-1}e_0,\\
 P_N^{-1}=\widetilde P_N^{-1}+
       \frac{u_Ne_0^\dagger\widetilde P_N^{-1}}
                 {1-e_0^\dagger u_N}.
 \end{gathered}
\end{equation}
The correction recovers the original $P_N^{-1}$. We check direct and adjoint solves against the original equations, using residuals relative to the applied source as well as backward residuals. Both the mass/source factor and the energy matrix enter these calculations.

Independent dense audits at $N=8,12$, $L_{\rm ang}=2$ rebuild the Gram from Chebyshev moments and polar quadrature. Gram entries agree to $7.71\times10^{-102}$ relatively, six energy-weighted singular residuals to $2.54\times10^{-60}$, and explicit mass--inverse identities to $6\times10^{-97}$. Three dense bordered-inverse comparisons at $260$ digits differ by at most $2.17\times10^{-98}$ relatively. At the $N=200$ pair midpoint, direct and updated-inverse singular residuals agree through $60$ reported digits. Two direct checks away from poles agree through $60$ digits at a broad bend and to $1.98\times10^{-59}$ relatively at a deep near-axis point. These are checks of finite-matrix arithmetic, not continuum error estimates.

A separate quadrature check increases working precision from $200$ to $240$ digits at $N=200$. The tested Gauss nodes and weights change by at most $6.6\times10^{-202}$ and $2.7\times10^{-201}$, respectively, and a factor--inverse identity has error $2.1\times10^{-201}$. At displacement $10^{-12}$ from the baseline positive $\ell=2$ gravitational-led pole, direct resolvent evaluation gives $s_E\kappa_E/|z-\omega_j|=1.000000000002262$. This checks the mutual consistency of the pole, energy, left vector and source map.

At $a=0$, independently discretized odd-parity RN master equations with the potentials of Refs.~\cite{Moncrief1974Odd,ZhouMaHuang2026} agree with the four-field electromagnetic $\ell=1$ and gravitational-led/electromagnetic-led $\ell=2$ fundamentals to $6.8\times10^{-10}$ at $Q=0.4$ and $1.2\times10^{-9}$ at $Q=0.8$ in binary64 $N=40$ tests. Continuation through $a=0,0.05,\ldots,0.3$, at $Q=0.4$, $m=1$, tracks all ten baseline families one-to-one. At $(a,Q,m)=(0.2,0.4,2)$, independent $N=96$, $L_{\rm ang}=8$ calculations differ from the printed separated frequencies in Ref.~\cite{Destounis2026} by $3.08\times10^{-10}$ and $1.72\times10^{-11}$ for the gravitational-led and electromagnetic-led fundamental modes. These comparisons support mode identification, not norm or resolvent convergence.

For the eight baseline $\ell=2,3$ fundamental modes, the largest changes are $2.53\times10^{-20}$ between $N=96$ and $200$ at $L_{\rm ang}=6$, and $6.15\times10^{-10}$ between $L_{\rm ang}=6$ and $8$ at $N=96$. These baseline frequency diagnostics do not bound angular errors of the finely tuned EP candidate. Appendix~\ref{app:binary64-checks} and Sec.~\ref{sec:resolution-tests} show that the candidate is not angularly converged at its selected pair-gap scale. A bordered characteristic function gives, at that unrounded candidate,
\begin{align}
 |F|&\simeq9.28\times10^{-146},&
 |F_\omega|&\simeq3.99\times10^{-143},\\
 F_{\omega\omega}&\simeq1068.42-2424.93i.
\end{align}
The auxiliary system remains nonsingular, while the original frequency equation has one independent eigenvector. The residual magnitudes depend on the normalization of $F$. They test the finite matrix and provide neither rigorous error bounds nor proof of a continuum EP.

\subsection{Independent binary64 checks}
\label{app:binary64-checks}

The resolution and truncation tests of Sec.~\ref{sec:resolution-tests} use a separate implementation of both discretized problems in IEEE binary64 arithmetic, built only from the equations in these appendices. For the scalar field, it discretizes the separated radial operator of Eqs.~\eqref{eq:scalar-polynomial-operators}--\eqref{eq:scalar-branch-pencil} on the nodes of Eq.~\eqref{eq:hypkn-radial-grid}. The spheroidal eigenvalue and its first two frequency derivatives are obtained from a spectral Legendre representation with $40$ odd levels ($\ell\le79$). Roots and double roots follow from Newton iteration on a bordered characteristic function, and the condition numbers from Eq.~\eqref{eq:branchkappa} with the energy of Eq.~\eqref{eq:hypkn-scalar-compact-energy}. This implementation reproduces the $n=0$ roots of Table~\ref{tab:revisionroots} to $3\times10^{-11}$, limited by the printed digits. The condition numbers agree to four or five significant figures: $8.1276\times10^4$, $1.7653\times10^4$ and $3.3616\times10^5$ for the positive fundamental at $N=200,160,240$, $3.0726\times10^5$ for the negative fundamental, and $5.0749\times10^7$ and $8.181\times10^9$ for QNM~1 and QNM~2 at Eq.~\eqref{eq:scalar-enhanced-background}. The double root of Eq.~\eqref{eq:scalar-ep-reference} is recovered to about $10^{-11}$ for $N\ge60$.

For GE, the implementation discretizes Eqs.~\eqref{eq:hypkn-GE-first-order-form}--\eqref{eq:hypkn-GE-potential-matrix} with the spin-weighted Jacobi bases of Eq.~\eqref{eq:hypkn-GE-angular-normalization}. It applies the Galerkin projection of $f_c(\mathcal D_m-izB)$ described above and uses Chebyshev--Lobatto collocation in $x$. The energy matrix is assembled from Eq.~\eqref{eq:ge-energy} by Gauss--Legendre quadrature, and condition numbers follow from Eq.~\eqref{eq:ge-gen-kappa}. All ten baseline frequencies of Table~\ref{tab:ge-coupled-modes} are reproduced to $6\times10^{-11}$ at $N=48$ and to $1.3\times10^{-11}$ at $N=64$. Omitting $f_c$ raises the $\ell=3$ discrepancies to $5.5\times10^{-10}$, so the projection prescription matters at this level. The baseline $\ell=2$ gravitational-led condition number is $6.849\times10^3$ at $N=20$, consistent with Fig.~\ref{fig:ge-stability}. At $t=10^{-3}$ and $N=100$, the pair gap is $8.28\times10^{-5}$, compared with $8.27879\times10^{-5}$ at $N=200$.

In binary64, roots near the pair carry roundoff of order $10^{-16}\kappa$, which makes direct GE pair roots unreliable for $N\gtrsim140$. The tests therefore use $\kappa d$, the comparison-mode condition numbers, and the following treatment of angular refinement. Let the matrix at truncation $L_1>L_0$ be partitioned into the functions retained at $L_0$ and those added. Then $P_{\rm eff}(z)=P_{00}(z)-P_{0x}(z)P_{xx}(z)^{-1}P_{x0}(z)$ has the same eigenvalues as the $L_1$ problem wherever $P_{xx}$ is invertible. Its block $P_{00}$ is the $L_0$ matrix. With right and left null vectors $r_i,l_i$ of $P_{00}(\omega_i)$,
\begin{equation}
 \begin{gathered}
 \delta\omega_i=-\frac{l_i^\dagger\,\delta P(\omega_i)\,r_i}{l_i^\dagger P_{00}'(\omega_i)\,r_i},\qquad
 \delta P=P_{\rm eff}-P_{00},\\
 \delta D=2(\omega_1-\omega_2)(\delta\omega_1-\delta\omega_2),
 \end{gathered}
 \label{eq:schur-discriminant}
\end{equation}
to first order. Because $D$ is analytic at the EP, $\delta D$ remains well conditioned even when the individual $\delta\omega_i$ are large compared with the gap. At $N=60$, Eq.~\eqref{eq:schur-discriminant} reproduces a direct $L_{\rm ang}=4\to6$ difference to $3\%$ for a change comparable to $D$ itself. At $N=100$, it agrees with direct $L_{\rm ang}=7,8,10$ calculations to within their seed-to-seed roundoff of about $3\times10^{-6}$ in $t_{\rm eff}$. These are checks of the finite matrices. As in the rest of this appendix, they provide no continuum error bounds for the condition numbers.

\subsection{Frequency interpolation and contour tracing}
\label{app:ge-interpolation}

The broad EP panel uses $433$ distinct measured frequencies, including the $197$ used in the local panel. A further $48$ frequencies, of which $24$ are local, are reserved for checks and excluded from interpolation. For selected poles at the same background, write
\begin{equation}\label{eq:ge-pole-factor}
 \begin{gathered}
 p_{\mathcal J}(z)=\prod_{j\in\mathcal J}(z-\omega_j),\\
 s_E(z)=\frac{|p_{\mathcal J}(z)|}
                  {\|p_{\mathcal J}(z)R_N(z)\|_E},\\
 h_{\mathcal J}(z)=\log_{10}s_E(z)-\log_{10}|p_{\mathcal J}(z)|.
 \end{gathered}
\end{equation}
Away from the poles, the scalar factor $p_{\mathcal J}$ can be taken out of the norm. We interpolate the measured remainder linearly on a triangular mesh, denoting this interpolation by $I_{\rm P1}$, and reconstruct
\begin{equation}\label{eq:ge-pole-interpolant}
 \widehat s_E(z)=|p_{\mathcal J}(z)|\,10^{I_{\rm P1}h_{\mathcal J}(z)}.
\end{equation}
Four pole factors are used for the wide panel and two for the local panel. The reference EP marked by a triangle occurs at a different background and does not enter the product. Every source value is evaluated from the full coupled resolvent, without inserted zero-valued measurements or a fitted residue model.

Closely spaced source frequencies are distinguished using exact integer tests of triangle orientation and circumcircles. These tests use the stored decimal coordinates after each axis has been scaled. Differences between source and check frequencies are evaluated at $300$ digits before conversion to binary64 distances and logarithms. Pole offsets and frame widths are also evaluated at high precision before rendering. The largest withheld discrepancy is $0.01601$ dex in the wide view. Local differences below $3.6\times10^{-15}$ dex reach the binary64 postprocessing floor and do not certify the underlying resolvent to that accuracy. These checks informed interpolation selection and are not independent uniform error bounds.

Colors, contours and validation queries all use the same interpolant. Contours are traced with uniform subdivision level $4$. Increasing this to level $5$ gives maximum sampled changes of $0.1813/1000$ and $0.1315/1000$ of the normalized frame. These are measured outside reference-marker disks of radius $0.008$ by the largest nearest-curve distance between the contour unions (Hausdorff distance). Subdivision adds no resolvent solves and does not bound the error in the interpolated field. The data include all numerical source values, precision settings and checks.
\section{Outgoing boundary conditions and energy weights}
\label{sec:domain}
\label{app:endpoints}

The energy \eqref{eq:ge-energy} measures the size of the four regularized fields. Finiteness of this energy is insufficient to enforce QNM boundary conditions, as the following endpoint behaviors illustrate. Near the horizon, use $y=1-x$ and the quantities defined in Eq.~\eqref{eq:hypkn-horizon-parameters}. An unwanted branch outgoing at the future horizon behaves as
\begin{equation}\label{eq:ge-horizon-indicial}
 Y_{\rm lower}\sim y^p,\qquad
 p=1+\frac{i(\omega-m\Omega_H)}{\kappa_H},
\end{equation}
Here ``upper'' and ``lower'' refer to components $(1,3)$ and $(2,4)$ of $Y$, respectively. The upper components are one power higher. At damped frequencies, $\Rew p=1-\Imw\omega/\kappa_H>1$, so the unwanted solution has finite energy and weighted radial derivative. Requiring square-integrable radial derivatives through second order ($H^2$ regularity) excludes its generic noninteger form at the baseline gravitational-led fundamental, where $\Rew p\simeq1.364$. For the near-EP pair, $\Rew p\simeq2.415$, so at least third-order ($H^3$) regularity is needed for the same local exclusion. These local tests depend on the field component and do not establish the full coupled QNM boundary-value problem.

For the scalar field, the unwanted incoming solution contains the factor $e^{-2izr_*}$ after the hyperboloidal rescaling. With $r=R/x$ and $\operatorname{Im}z<0$, its leading exponential decays faster than any power of $x$. The coupled fields have the analogous behavior: at future null infinity the unwanted incoming branch behaves as
\begin{equation}\label{eq:ge-infinity-flat}
 \begin{gathered}
 Y_{\rm upper}\sim x^{-4+4i\omega}\exp(-2iR\omega/x),\\
 Y_{\rm lower}=O(x^2Y_{\rm upper}).
 \end{gathered}
\end{equation}
When $\Imw\omega<0$, this solution and all its derivatives vanish at $x=0$. It passes every test based on square-integrability of finitely many radial derivatives, even when extended by zero beyond the endpoint. Selecting the outgoing solution requires a stronger condition. Analyses of scalar Kerr fields impose bounds on derivative growth, known as Gevrey regularity \cite{GajicWarnick2024}. A corresponding result has yet to be established for the coupled KN fields. Finite polynomial approximations impose stronger endpoint regularity than the energy norm controls. Sec.~\ref{sec:discussion} discusses the consequences for relating finite-matrix tests to a continuum response.

The relative field weight is also part of the perturbation model. Replacing the connection-pair contribution by $\lambda>0$ times its value changes equivalent induced norms by at most
\begin{equation}
 \sqrt{\frac{\max(1,\lambda)}{\min(1,\lambda)}}.
 \label{eq:ge-weight-bound}
\end{equation}
This factor also bounds changes in the resolvent norm, threshold and simple-mode condition number. Choosing a different weight for each mode would alter the comparison itself.

At real frequencies, Ref.~\cite{Motohashi2026} extracts the Maxwell amplitude by subtracting a gravitational contribution from a mixed radiative invariant. This reconstruction is separate from the energy balance in Eq.~\eqref{eq:ge-balance}. The restrictions on physically admissible sources are discussed in Sec.~\ref{sec:discussion}.

\bibliography{biblio}
\end{document}